\documentclass[aps,superscriptaddress,amsmath,amssymb,floatfix,twocolumn,showpacs,amsfonts,longbibliography]{revtex4-1}
\usepackage{times}
\usepackage[varg]{txfonts}
\usepackage{textcomp}
\usepackage{graphicx}
\usepackage{subfigure}
\usepackage{tabu}
\usepackage{color}
\usepackage[colorlinks=true,citecolor=blue,urlcolor=blue,linkcolor=blue,hyperindex]{hyperref}
\usepackage{braket}
\usepackage{overpic}
\usepackage{amssymb}
\allowdisplaybreaks

\begin{document}
\title{Klein tunneling of Weyl magnons}
\author{Chengkang Zhou}
\author{Luyang Wang}
\email{wangly39@mail.sysu.edu.cn}
\author{Dao-Xin Yao}
\email{yaodaox@mail.sysu.edu.cn}
\affiliation{State Key Laboratory of Optoelectronic Material and Technologies, School of Physics, Sun Yat-sen University, Guangzhou 510275, China}

\date{\today}

\begin{abstract}
Similar to Weyl semimetals, in magnetic materials, magnon bands can host Weyl points, around which the bosonic excitations are called Weyl magnons.
Here, we investigate the Klein tunneling of Weyl magnons, during the process of which Weyl magnons tunnel through a high potential barrier. Specifically, we study the magnetization current carried by Weyl magnons in a quasi-one-dimensional magnetic wire, in the middle of which a gate magnetic field is applied to generate a potential barrier. The transmission probability is calculated and the Landauer-B\"uttiker formalism is used to find the magnetization current. Various types of Weyl magnons are considered, including isotropic, tilted, and double Weyl magnons. Unlike in Weyl semimetals where fermionic statistics is in charge and the current oscillates with the gate field as a result of Fabry-P\'erot resonances, here Bose distribution smears out the oscillations. We find that the tilting of the Weyl cone causes the decrease of the magnetization current from Klein tunneling, while for double Weyl magnons, Klein tunneling is absent in the direction of quadratic dispersion, but is enhanced in the direction of linear dispersion. Our results show that the behaviour of the current-voltage characteristics of magnons is rather different from that of electrons due to different statistics, although the single-particle properties, such as the transmission probabilities, are the same for both Weyl bosons and Weyl fermions.

\end{abstract}
\maketitle

\section{Introduction}
\label{INTRODUCTION}
The Klein tunneling effect is one of the most interesting relativistic quantum effects. It is also known as the Klein paradox, which was born ninety years ago, but has only been observed recently in condensed matters\cite{rose1961relativistic,dombey1999seventy,calogeracos1999history,holstein1998kleins,katsnelson2006chiral,allain2011klein}. The early papers investigated the scattering of a relativistic particle off a high potential step in the context of the Dirac equation and discussed the paradox first observed by Klein. The so-called Klein tunneling effect refers to a process during which an incoming relativistic particle tunnel into a potential step when its height $V$ exceeds twice the particle's rest energy $mc^2$ (where $m$ is the mass and $c$ is the speed of light). The transmission probability $T$ in the limit of a high potential step barely depends on the step's height and the backscattering is absent, which means the step keeps perfect transparency. In stark contrast, the transmission probability $T$ in non-relativistic tunneling decays exponentially as $V$ increases. Such a difference is mainly caused by the essential property of the Dirac equation that the particle states with positive and negative energy are intimately connected with each other. In fact, they are described by different components of the same spinor wave function, which is known as the charge-conjugation symmetry. Additionally, for the case of oblique incidence, a potential barrier plays the role of the Fabry-P\'erot interferometer for particle wave functions, which leads to the high transmission probability at special incident angles\cite{allain2011klein,zhang2018oblique,huang2018klein,yesilyurt2016klein,andrei2008klein}.

Although Klein tunneling has been well understood theoretically, it is hard to be observed experimentally in particle physics because of the requirement of an extremely large electric field for massive particles. The discovery of graphene has provided a platform to explore relativistic quantum effects in condensed matter systems\cite{Neto2009RMP}. The quasiparticles in graphene are massless Dirac fermions, suggesting that the Klein tunneling effect is observable in this system\cite{katsnelson2006chiral}. The experimental studies of the electronic transport in graphene heterostructures opens the door to the observation of Klein tunneling\cite{Sonin2009EffectKleintunneling,young2009quantum,rossi2010signatures,young2011electronic}. Combined with Fabry-P\'erot resonances, Klein tunneling yields oscillations in the current-voltage curve in these graphene devices.

Derivatives of Klein tunneling in systems beyond Dirac particles have been suggested and studied. For example, pseudospin-1 particles show omnidirectional Klein tunneling with $T=1$ when their energy equals half the barrier height\cite{urban2011barrier,xu2014omnidirectional}. In a lattice with coexisting pseudospin-1 and Dirac particles, pseudospin-nonconserving Klein tunneling can occur, during which the pseudospin of the tunneling particle is not conserved\cite{wang2018coexistence}. 

The discovery of three-dimensional Dirac and Weyl semimetals\cite{Armitage2018RMP} provides new platforms to study relativistic particles. Novel physical effects have been studied in these materials, such as Fermi arc surface states\cite{Wan2011PRB}, anomalous Hall effect\cite{Yang2011PRB}, chiral anomaly\cite{Son2013PRB}, three-dimensional quantum Hall effect\cite{Wang2017PRL,Zhang2019Nature} and Imbert-Fedorov shift\cite{Jiang2015PRL,Yang2015PRL,Wang2017PRB,Hao2019PRB}. Klein tunneling of Weyl fermions has also been investigated\cite{o'brien2016magnetic,yesilyurt2016klein,Bai2016,Nguyen2018Klein}. It is an attractive topic how the transmission probability evolves with the Weyl cone changing from Type-I to Type-II\cite{o'brien2016magnetic}, since the tilting of a Weyl cone shows a strong resemblance to the behavior of a light cone of a probe particle approaching a black hole horizon\cite{hills2017current-voltage,novikov1963light}. Although the transmission properties are mostly discussed in the context of fermions, these single- particle properties are also applicable to bosonic systems, such as photonic crystals with Weyl points\cite{lu2013weyl,lu2015experimental,wang2016topological} and magnonic Weyl materials.

The research interests in topological magnons are increasing in recent years. Magnons are quasiparticles of collective spin excitations in magnetic systems, which emerge from the standard linearized Holstein-Primakoff transformation. Weyl magnons, the topological magnons which obey Weyl equation within a certain energy range, have been predicted to exist in pyrochlores\cite{li2016weyl, Mook2016Tunable, Owerre2018WeylMagnons, Shaokai2018WeylMagnonsPyrochlore}. Topological properties of Weyl magnons have been discussed, such as arc surface states and chiral anomaly\cite{Su2017MagnonicWeyl, Su2017ChiralAnomaly}.

In this paper, we focus on the Klein tunneling effect of Weyl magnons and discuss the magnon current influenced by this effect. The moving magnons cause a spin current, which can be seen as a magnetization current as well\cite{Loss2003MagnetizationTransport,nakata2017spin}. We consider a quasi-one-dimensional magnetic wire in which the magnetization current is carried by Weyl magnons. By setting a constant gate magnetic field in the middle of the wire, we study the Klein tunneling effect and the transport properties of isotropic, tilted and double Weyl magnons. Finally, with these transmission properties, we present the curves of the magnetization current against the potential barrier height. Since the ballistic magnon transport has been achieved in $\text{Nd}_2\text{Cu}\text{O}_4$ at low temperature\cite{li2005ballistic}, and the magnon transport has also been observed in $\text{YIG}$\cite{ganzhorn2016magnon,serga2010yig}, we expect that our results are instructional to future transport experiments of Weyl magnons.

The paper is organized as follows. In Sec.\ref{Sec:MODEL}, we propose an experimental setup to study the Klein tunneling of Weyl magnons, in which Weyl magnons tunnel through a potential barrier. In Sec.\ref{Sec:SPTRPBarrier}, the transmission probabilities of various types of Weyl magnons are calculated. In Sec.\ref{Sec:MConductance}, we describe the ballistic magnon transport using the Landauer-B\"uttiker approach. The magnetization current carried by Weyl magnons is calculated, and the influences of Klein tunneling and Fabry-P\'erot resonances are discussed. In Sec.\ref{Sec:CONCLUSION}, we conclude our results.

\section{Weyl magnons and ballistic magnon transport}
\label{Sec:MODEL}
Magnons are the bosonic quasiparticles of collective spin-wave excitations in magnetic spin systems where spins precess coherently around the direction of the local magnetic order. They are usually studied with the Holstein-Primakoff transformation, which transforms spin operators to annihilation and creation operators of magnons. In momentum space, the Hamiltonian of magnons in a two-band model is given by $\mathcal{H}=\sum_{\bf k}\Psi_{\bf k}^\dagger H_{m}({\bf k})\Psi_{\bf k}$, in which $\Psi_{\bf k}^\dagger=(a^{\dagger}_{\bf k},b^{\dagger}_{\bf k})$ refers to the magnon creation operator. This Hamiltonian can be viewed as the bosonic tight binding model and the topological band theory can be applied to investigate magnon bands. As a result, the concept of Dirac and Weyl materials has been extended from electronic systems to magnonic systems. Near the Weyl points, the effective Hamiltonian of magnons is the Weyl Hamiltonian,
\begin{equation}
	\begin{aligned}
		H_{m}({\bf k})=E_{0}\sigma_{0}+\hbar\sum_{a=x,y,z}v_{a}k_{a}\sigma_{a},
	\end{aligned}
\end{equation}
where $E_{0}$ is the Weyl point energy, $\sigma_{0}$ is the $2\times2$ identity matrix, $k_a$ and $v_{a}$ refer to the momentum deviation from the Weyl point and the velocity parameter in the $a$-direction, respectively, and $\sigma_{a}$'s are the Pauli matrices. The magnons have a linear dispersion given by
\begin{equation}
	\begin{aligned}
		E_{m}({\bf k})=E_{0}\pm\hbar \sqrt{\sum_{a=x,y,z}v_{a}^2k_{a}^2}.
	\end{aligned}
\end{equation}
In fermionic systems, the linear dispersion can lead to Klein tunneling\cite{yesilyurt2016klein,Bai2016,Nguyen2018Klein}, similar to the Klein tunneling of Dirac fermions in graphene heterostructures\cite{katsnelson2006chiral,Sonin2009EffectKleintunneling,young2009quantum,rossi2010signatures,young2011electronic}, as well as behaviors analogous to light rays in optical media, such as refraction, reflection and Fabry-P\'erot resonances\cite{allain2011klein, Nguyen2018Klein}. Such effects can also happen in bosonic systems. Inspired by the experimental and theoretical developments of the magnon  transport\cite{Loss2003MagnetizationTransport, Nakata2017MagnonicHall, nakata2017spin}, we investigate Klein tunneling and related effects of Weyl magnons in magnetic systems.

\begin{figure}[ht]
	\centering
	{\includegraphics[height=5cm]{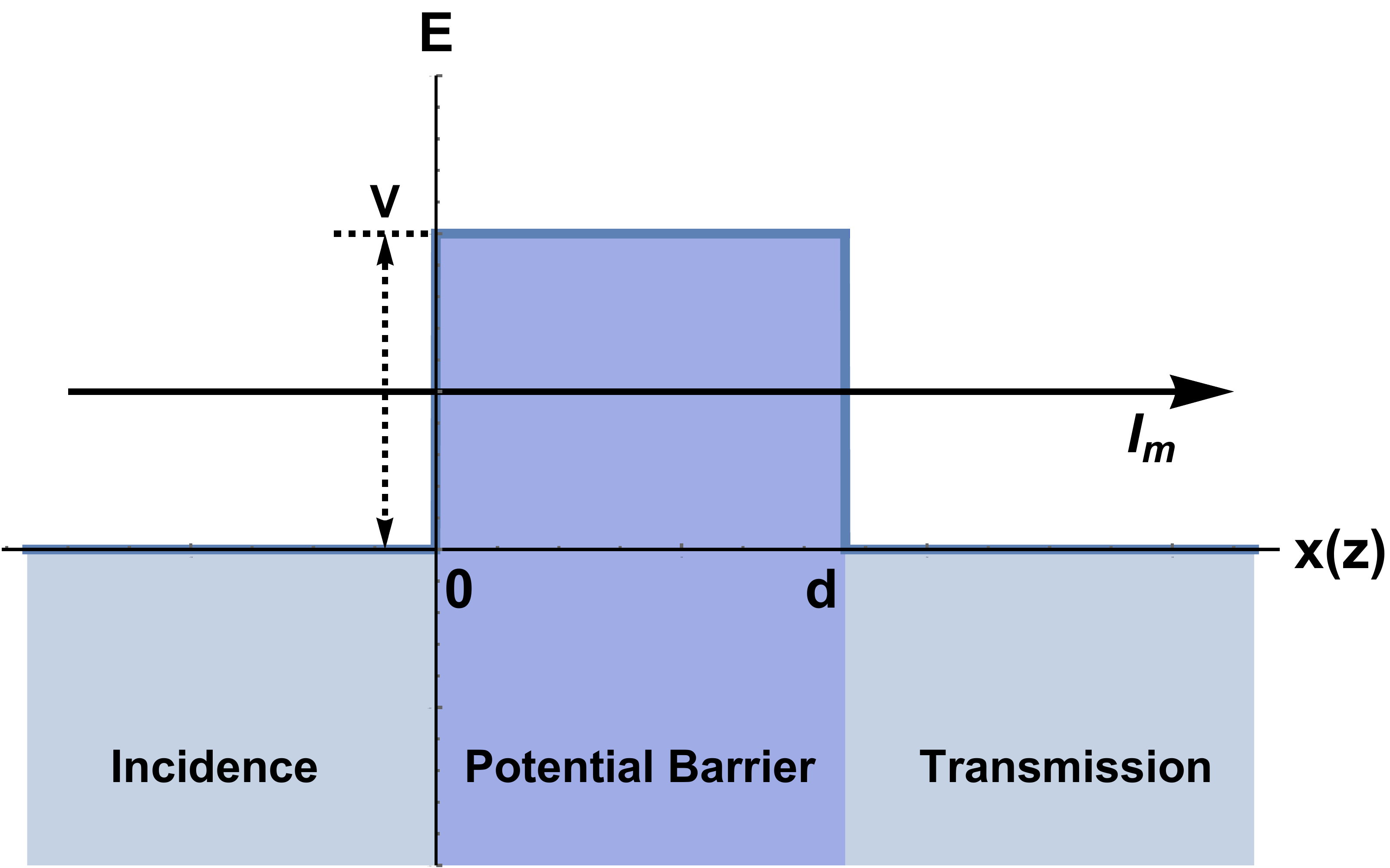}}
	\caption{Schematic diagram of the tunneling model of Weyl magnons. A magnetization current carried by Weyl magnons flows in a quasi-one-dimensional magnetic wire along the $x$- or $z$-direction\cite{Su2017MagnonicWeyl}. A potential barrier with height $V=\textsl{g}\mu_{B}B_{g}$ and width $d$ is placed in the middle of the wire, which can be achieved by a gate magnetic field. The incident, potential barrier, and transmission regions are marked. We also consider the case that $V$ is negative, which becomes a potential well problem.}
	\label{Fig01}
\end{figure}

We consider the ballistic magnon transport in a quasi-one-dimensional magnetic wire influenced by a potential barrier or well, as shown in Fig.\ref{Fig01}. Driven by the difference of magnon chemical potentials, magnons are transported through the wire in the $x$-or $z$-direction. A potential barrier with height $V=\textsl{g}\mu_{B}B_{g}$ and width $d$ is placed in the middle of the wire, where $\textsl{g}$ is the Land\'e $g$-factor, $\mu_B$ the Bohr magneton and $B_{g}$  a constant gate magnetic field. When $V$ is negative, it becomes a potential well. In the following discussions, when we mention the potential barrier $V$, it may also refer to a potential well when $V<0$. We put a cutoff around the Weyl points in the magnon bands and focus on the tunneling effect and the transport of Weyl magnons within the cutoff. The Landauer-B\"uttiker approach is applied to calculate the magnetization current carried by magnons in this system\cite{Loss2003MagnetizationTransport, nakata2017spin, zhang2013topological}.

\section{Transmission through a rectangular potential barrier}
\label{Sec:SPTRPBarrier}

\subsection{Isotropic Weyl magnons}
\label{SubSec:WMagnon}

Firstly, we focus on the isotropic Weyl cone in magnon bands. In this case, the magnons near the Weyl points obey the original Weyl equation. Rescale $v_{a}k_{a}\rightarrow k_{a}$ and let $\hbar=1$, the effective Hamiltonian of isotropic Weyl magnons can be written as
\begin{equation}
	\begin{aligned}
		H_{W}({\bf k}) & =\sum_{a=x,y,z}k_{a}\sigma_{a}+V(x)\sigma_{0},
	\end{aligned}
\end{equation}
where $V(x)$ is a rectangular potential barrier given by
\begin{equation}
	V(x)=\left\{
	\begin{aligned}
		V & \qquad 0\!<\!x\!<\!d,                \\
		0 & \qquad x\!<\!0\quad \mbox{or}\quad x\!>\!d.
	\end{aligned}
	\right.
	\label{barrier}
\end{equation}
which is introduced by a constant gate magnetic field. The eigenenergy of $H_W({\bf k})$ is
\begin{equation}
	\begin{aligned}
		E_{W}({\bf k})=sk+V(x),
	\end{aligned}
\end{equation}
where $k=\sqrt{k_x^2+k_y^2+k_z^2}$ and $s=\pm1$ is the magnon band index. When magnon transport happens in the $x$-direction, the wave function of the incident magnons with energy $\varepsilon$ is given by
\begin{equation}
	\begin{aligned}
		\psi(x) & =
		\frac{1}{\sqrt{A}}\left[
			\begin{array}{c}
				k_z+\varepsilon \\
				k_x+ik_y
			\end{array}
			\right]e^{ik_xx}
            \equiv\left[
			\begin{array}{c}
				\psi_{1}(k_x) \\
				\psi_{2}(k_x)
			\end{array}
			\right]e^{ik_xx},
	\end{aligned}
	\label{wfiw}
\end{equation}
in which $A=2\varepsilon(\varepsilon+k_z)$. Since $k_y$ and $k_z$ are conserved during the process, they are regarded as fixed parameters, and the plane wave part $e^{i(k_yy+k_zz)}$ is omitted.

To solve the tunneling problem, we assume that the incident magnon propagates with momentum ${\bf k}$. Note that the propagating direction is determined by the group velocity ${\bf v}=dE/d{\bf k}$. The magnon wave functions in the incident region and the transmission region are, respectively,
\begin{equation}
	\begin{aligned}
		\psi^{\mathrm{i}}(x)=\left[
			\begin{array}{c}
				\psi_{1}(k_x) \\
				\psi_{2}(k_x)
			\end{array}
			\right]e^{ik_xx}+r\left[
			\begin{array}{c}
				\psi_{1}(-k_x) \\
				\psi_{2}(-k_x)
			\end{array}
			\right]e^{-ik_xx},
	\end{aligned}
	\label{WaveFIn}
\end{equation}
and
\begin{equation}
	\begin{aligned}
		\psi^{\mathrm{t}}(x)=t\left[
			\begin{array}{c}
				\psi_{1}(k_x) \\
				\psi_{2}(k_x)
			\end{array}
			\right]e^{ik_xx},
	\end{aligned}
	\label{WaveFTr}
\end{equation}
where $t$ and $r$ are the coefficient of transmission and reflection, respectively. Inside the potential barrier, the wave function is
\begin{equation}
	\begin{aligned}
		\psi^{\mathrm{p}}(x)=a\left[
			\begin{array}{c}
				\psi_{1}(q_x) \\
				\psi_{2}(q_x)
			\end{array}
			\right]e^{iq_xx}+b\left[
			\begin{array}{c}
				\psi_{1}(-q_x) \\
				\psi_{2}(-q_x)
			\end{array}
			\right]e^{-iq_xx},
	\end{aligned}
	\label{WaveFPb}
\end{equation}
in which $q_x=\sqrt{(\varepsilon-V)^2-k_y^2-k_z^2}$.

From the continuity of wave functions, we have
\begin{equation}
	\begin{aligned}
		\psi^{\mathrm{i}}(0) & =\psi^{\mathrm{p}}(0),  \\
		\psi^{\mathrm{p}}(d) & =\psi^{\mathrm{t}}(d),
	\end{aligned}
	\label{continuity}
\end{equation}
from which we can solve the transmission probability $T=|t|^2$. To present the relation between $T$ and the momentum of the incident magnons, we introduce the spherical coordinates $(k,\theta,\phi)$ to the momentum space, which are related to the Cartesian coordinates by
\begin{equation}
	\begin{aligned}
		k_x & =k\sin\theta\cos\phi, \\
		k_y & =k\sin\theta\sin\phi, \\
		k_z & =k\cos\theta.
	\end{aligned}
	\label{SphericalCoordinates}
\end{equation}
\begin{figure}[t]
	\centering
	\subfigure[]{\includegraphics[height=3.5cm]{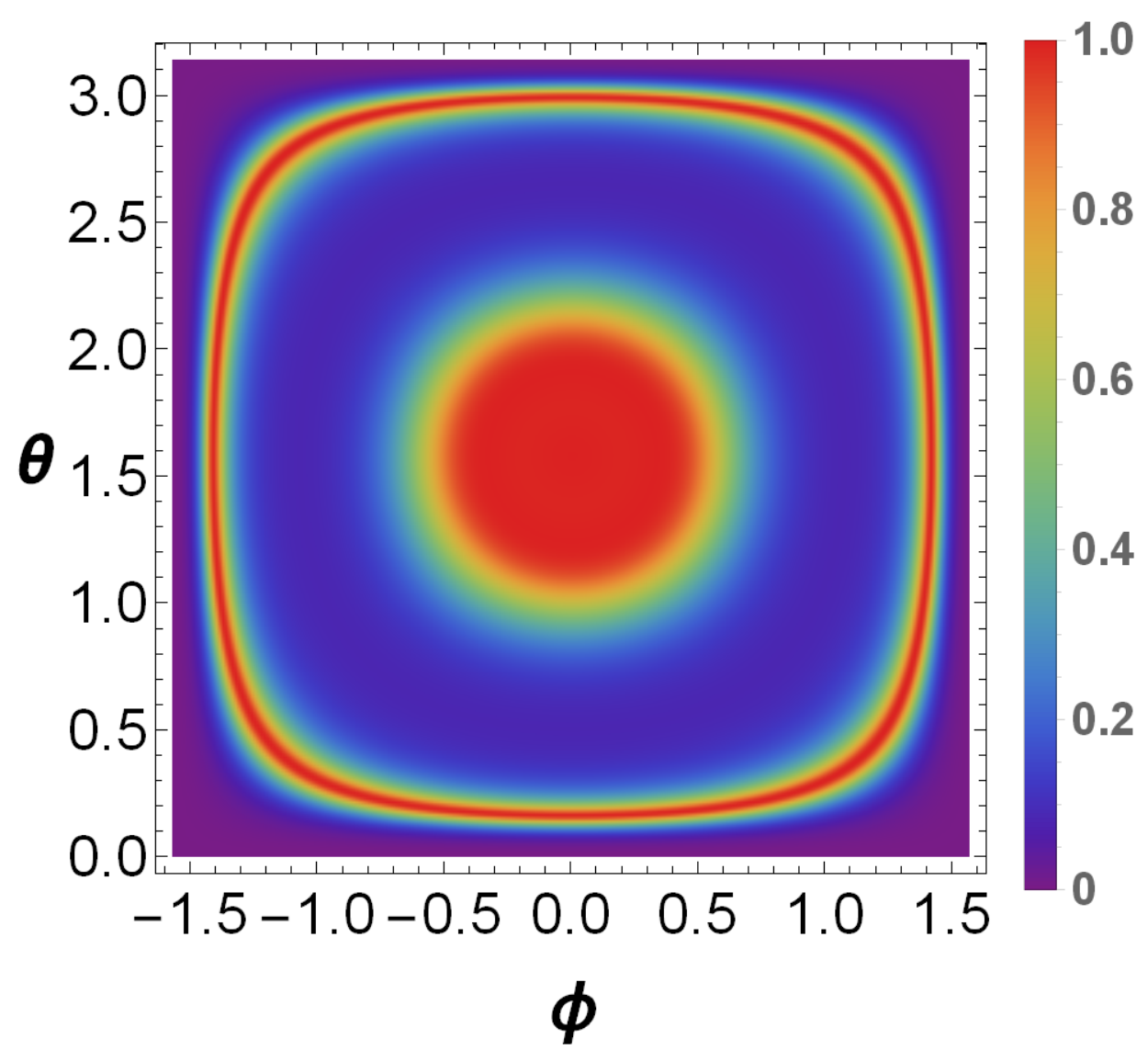}\label{Fig02a}}
	\subfigure[]{\includegraphics[height=3.5cm]{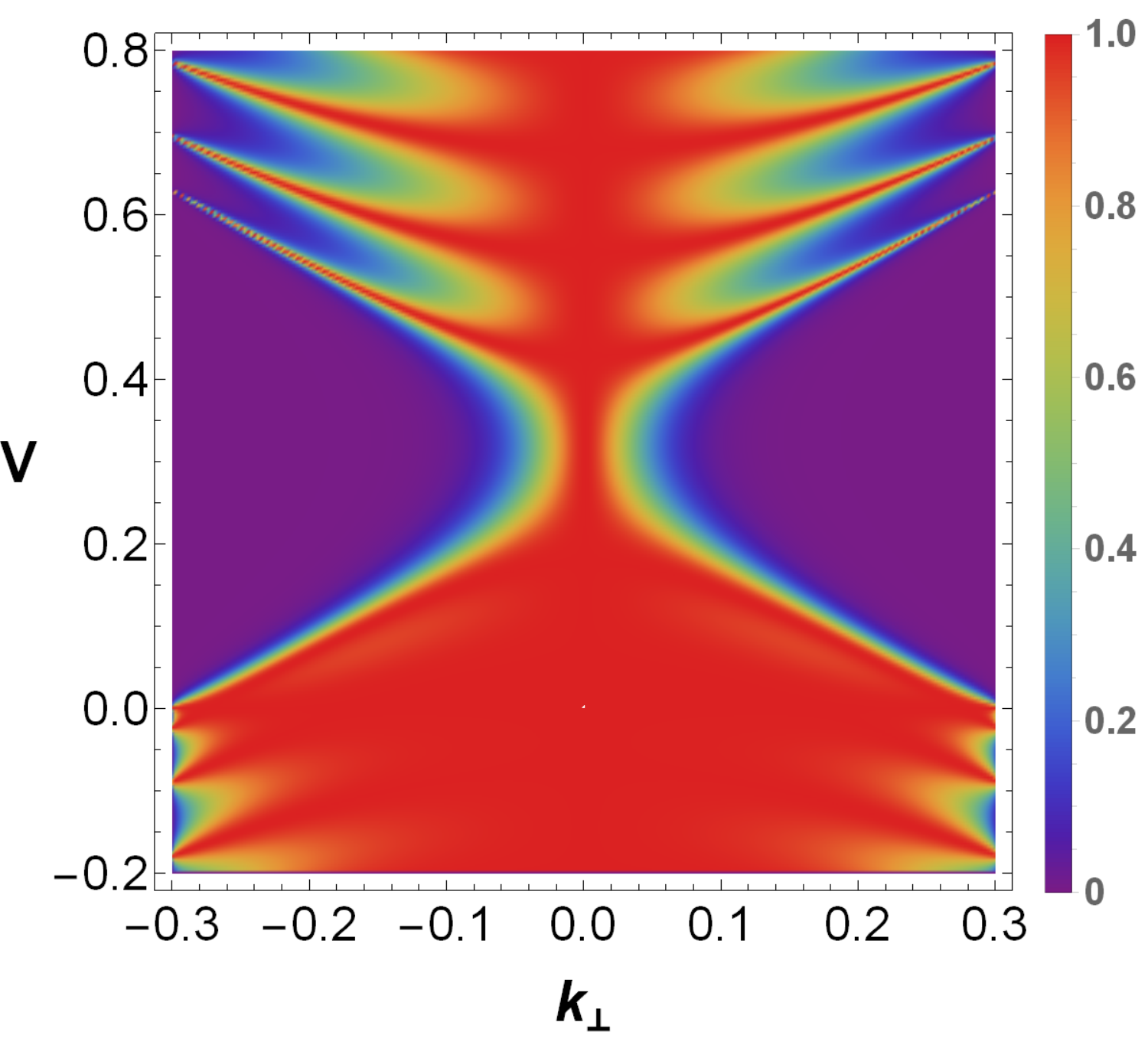}\label{Fig02b}}
	\caption{The dependence of the transmission probability of isotropic Weyl magnons on (a) the momentum angle, with $k=0.3$, $V=0.6875$, and $d=8\pi$, and (b) the barrier height, with $k=0.3$ and $d=8\pi$, and $k_{\perp}=\pm \sqrt{k_y^2+k_z^2}$.}
	\label{Fig02}
\end{figure}

The transmission probability $T$ is shown in Fig.\ref{Fig02a} and Fig.\ref{Fig02b} corresponding to the dependence on the momentum angles $(\theta, \phi)$ and on the barrier height $V$, respectively. Since magnons obey the Weyl equation, they have the same tunneling properties as Weyl fermions in the electronic case. The tunneling exhibits two well-known phenomena. First, $T=1$ independent of $V$ when $k_y=k_z=0$, which is known as the Klein tunneling effect. Second, resonant peaks appear in oblique incident directions, which is called Fabry-P\'erot resonances, with the resonance condition
\begin{equation}
	\begin{aligned}
    q_x d=n\pi, \ n=0, \pm1,\pm2...
	\end{aligned}
	\label{ResonanceConditions}
\end{equation}

When the tunneling happens in the $z$-direction, since the Weyl cone is isotropic, there is no difference in the transmission properties.
%

\subsection{Weyl magnons with tilted energy dispersion}
\label{SubSec:TWMagnon}
When the Weyl cone is tilted, the tunneling properties change. We assume that the tilting appears in the $x$-direction and study magnon transport happening in both the $x$- and $z$-directions. We consider the former first. The effective Hamiltonian can be expressed as
\begin{equation}
	\begin{aligned}
		H_{TW}({\bf k}) & =\sum_{a=x,y,z}k_{a}\sigma_{a}+[uk_x+V(x)]\sigma_{0} \\
		          & =\left(
		\begin{array}{cc}
				u k_x+k_z+V(x) & k_x-i k_y     \\
				k_x+i k_y      & uk_x-k_z+V(x) \\
			\end{array}
		\right)
	\end{aligned}
	\label{TWHam}
\end{equation}
in which $V(x)$ is given by Eq.(\ref{barrier}) and $u$ describes the tilting in the $x$-direction. We will assume $|u|<1$ in the following discussions. The eigenenergy is
\begin{equation}
	\begin{aligned}
		E_{TW}(k)=u k_x+sk+V(x),
	\end{aligned}
\end{equation}
where $s=\pm1$. The constant energy surface is given by
\begin{equation}
	\begin{aligned}
		\frac{\varepsilon^2}{1-u^2}=(1-u^2)(k_x+\frac{\varepsilon u}{1-u^2})^2+k_y^2+k_z^2,
	\end{aligned}
	\label{EnergyS}
\end{equation}
where $\varepsilon$ refers to the energy of incident magnons. Since $uk_x\sigma_{0}$ is diagonal, the Hamiltonian shares the same eigenvector with the untilted case, and the wave functions also take the form of Eqs.(\ref{WaveFIn}), (\ref{WaveFTr}) and (\ref{WaveFPb}). However, the wave vectors change due to the tilting of the dispersion. In the incident region, the wave vector of the reflected magnon becomes
\begin{equation}
	\begin{aligned}
		k'_{x} & =-\frac{2\varepsilon u}{1-u^2}-k_x.
	\end{aligned}
	\label{k'_x}
\end{equation}
In the barrier region, the wave vectors are
\begin{equation}
	\begin{aligned}
		q_{x,s} & =\frac{-(\varepsilon-V)u+s\sqrt{(\varepsilon-V)^2-(1-u^2)(k_y^2+k_z^2)}}{1-u^2}.
	\end{aligned}
\end{equation}
With the group velocity given by $v_x=dE/dk_x$, it is easy to see that $q_{x,+}$ describes the magnon wave propagating from $x=0$ to $x=d$ and $q_{x,-}$ is associated with the wave propagating in the opposite direction. Moreover, $(\varepsilon-V)^2<(1-u^2)(k_y^2+k_z^2)$ tells the low transmission area. Finally, the wave functions become
\begin{equation}
	\begin{aligned}
		\psi^{\mathrm{i}}(x) & =\left[
			\begin{array}{c}
				\psi_{1}(k_x) \\
				\psi_{2}(k_x)
			\end{array}
			\right]e^{ik_xx}+r\left[
			\begin{array}{c}
				\psi_{1}(k'_x) \\
				\psi_{2}(k'_x)
			\end{array}
			\right]e^{ik'_xx},              \\
		\psi^{\mathrm{p}}(x) & =a\left[
			\begin{array}{c}
				\psi_{1}(q_{x,+}) \\
				\psi_{2}(q_{x,+})
			\end{array}
			\right]e^{iq_{x,+}x}+b\left[
			\begin{array}{c}
				\psi_{1}(q_{x,-}) \\
				\psi_{2}(q_{x,-})
			\end{array}
			\right]e^{iq_{x,-}x},           \\
		\psi^{\mathrm{t}}(x) & =t\left[
			\begin{array}{c}
				\psi_{1}(k_x) \\
				\psi_{2}(k_x)
			\end{array}
			\right]e^{ik_xx}.
	\end{aligned}
\end{equation}
The transmission probability $T=|t|^2$ is calculated from the continuity of wave functions, as shown in Fig.\ref{Fig04}. Note that all incident magnons satisfy
\begin{equation}
	\begin{aligned}
		\frac{\varepsilon^2}{1-u^2}>k_y^2+k_z^2.
	\end{aligned}
	\label{ConditionKTWX}
\end{equation}
Again we use $k_{\perp}^2=k_y^2+k_z^2$ and $\gamma=\arcsin(k_y/k_{\perp})$ for a better presentation in Fig.\ref{Fig04}.

\begin{figure}
	\centering
    \subfigure[]{\includegraphics[height=3.5cm]{Fig02a.pdf}\label{Fig03a}}
	\subfigure[]{\includegraphics[height=3.5cm]{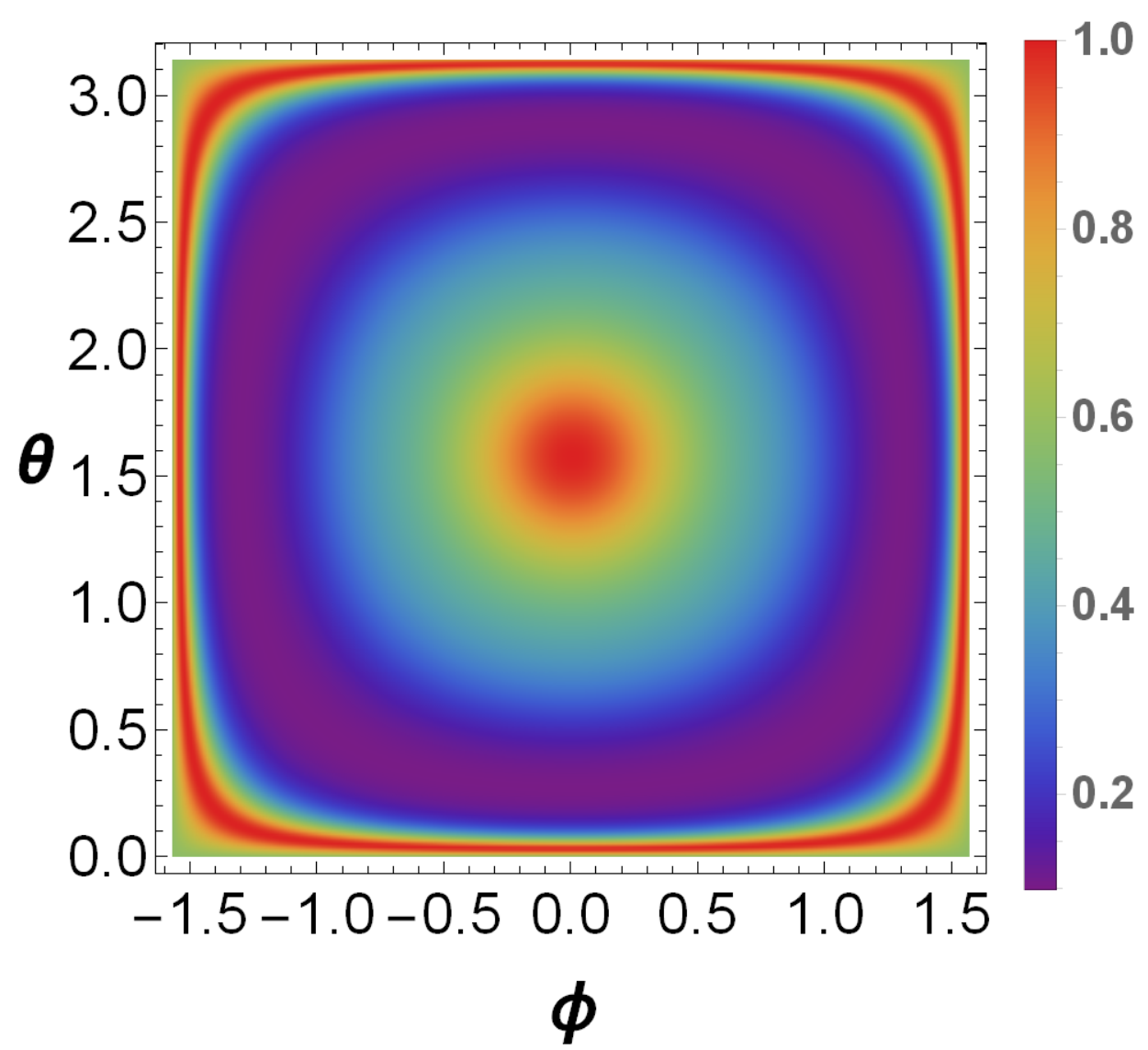}\label{Fig03b}}
	\subfigure[]{\includegraphics[height=3.5cm]{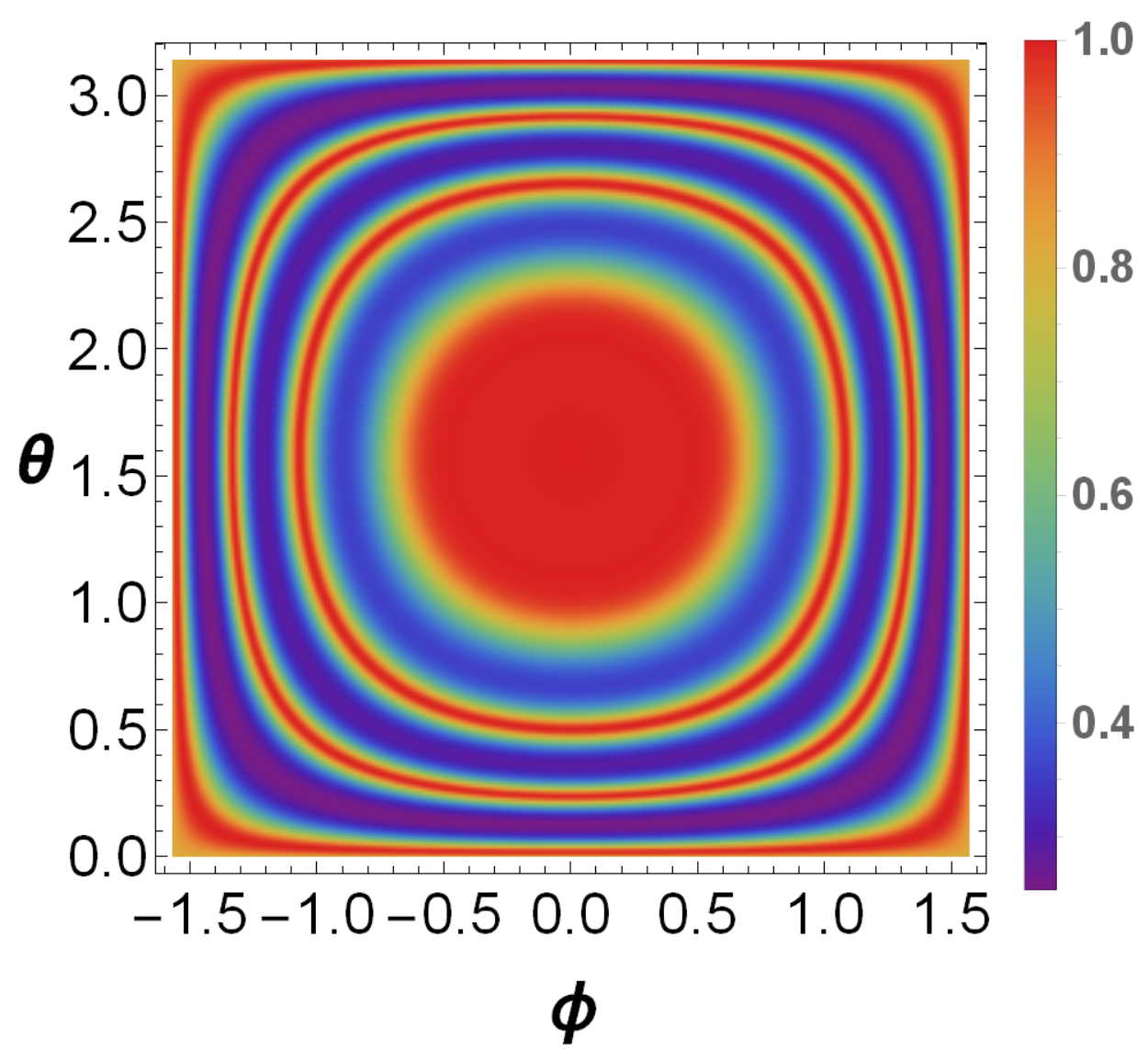}\label{Fig03c}}
	\subfigure[]{\includegraphics[height=3.5cm]{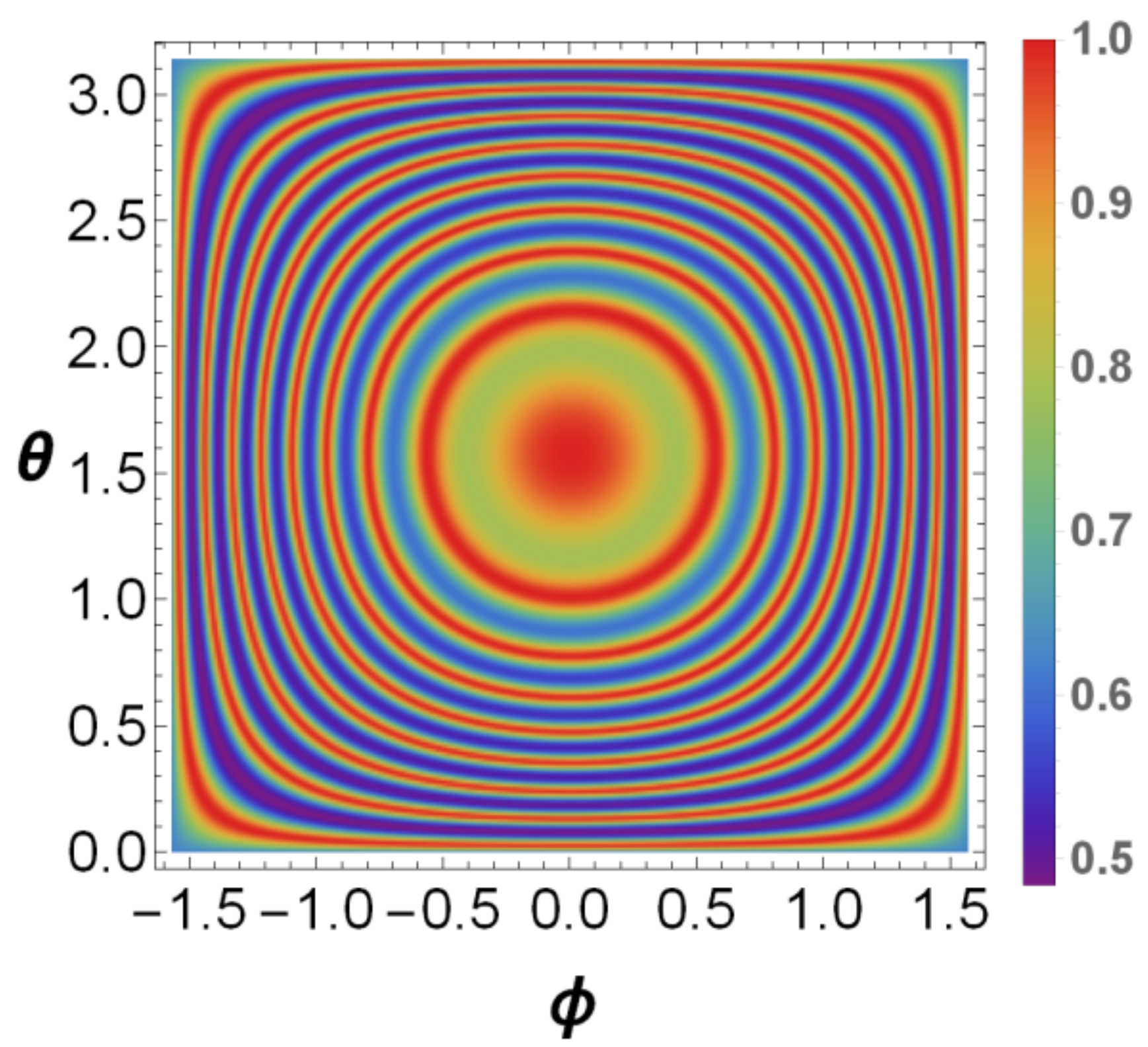}\label{Fig03d}}
	\caption{The momentum angle dependence of the transmission probability for (a) $u=0$, (b) $u=\frac{1}{2}$, (c) $u=\frac{3}{4}$, and (d) $u=\frac{7}{8}$. The fixed parameters are $k=0.3$, $V=0.6875$, and $d=8\pi$.}
	\label{Fig03}
\end{figure}

\begin{figure}
	\centering
	\subfigure[]{\includegraphics[height=3.5cm]{Fig02b.pdf}\label{Fig04a}}
	\subfigure[]{\includegraphics[height=3.5cm]{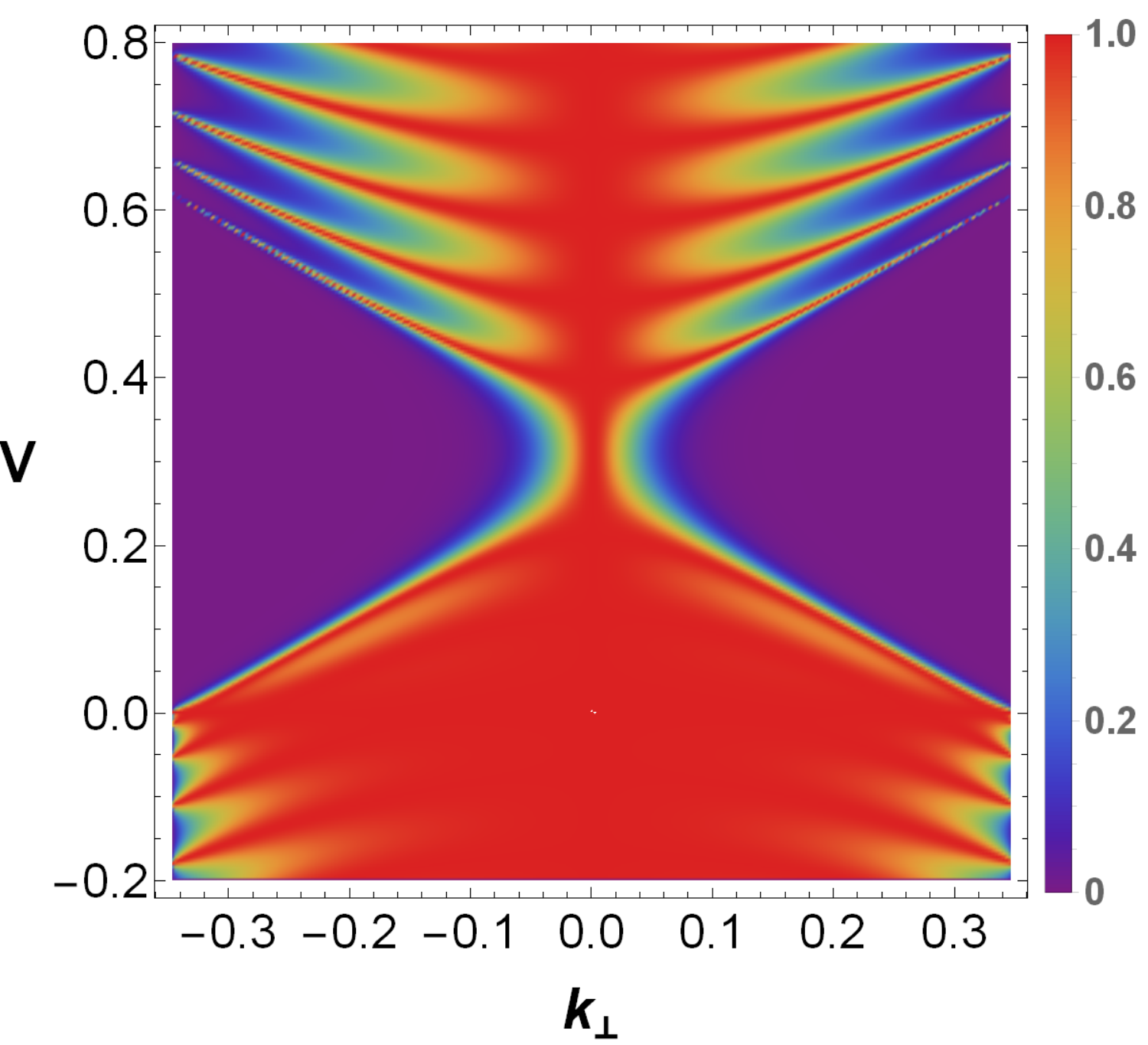}\label{Fig04b}}
	\subfigure[]{\includegraphics[height=3.5cm]{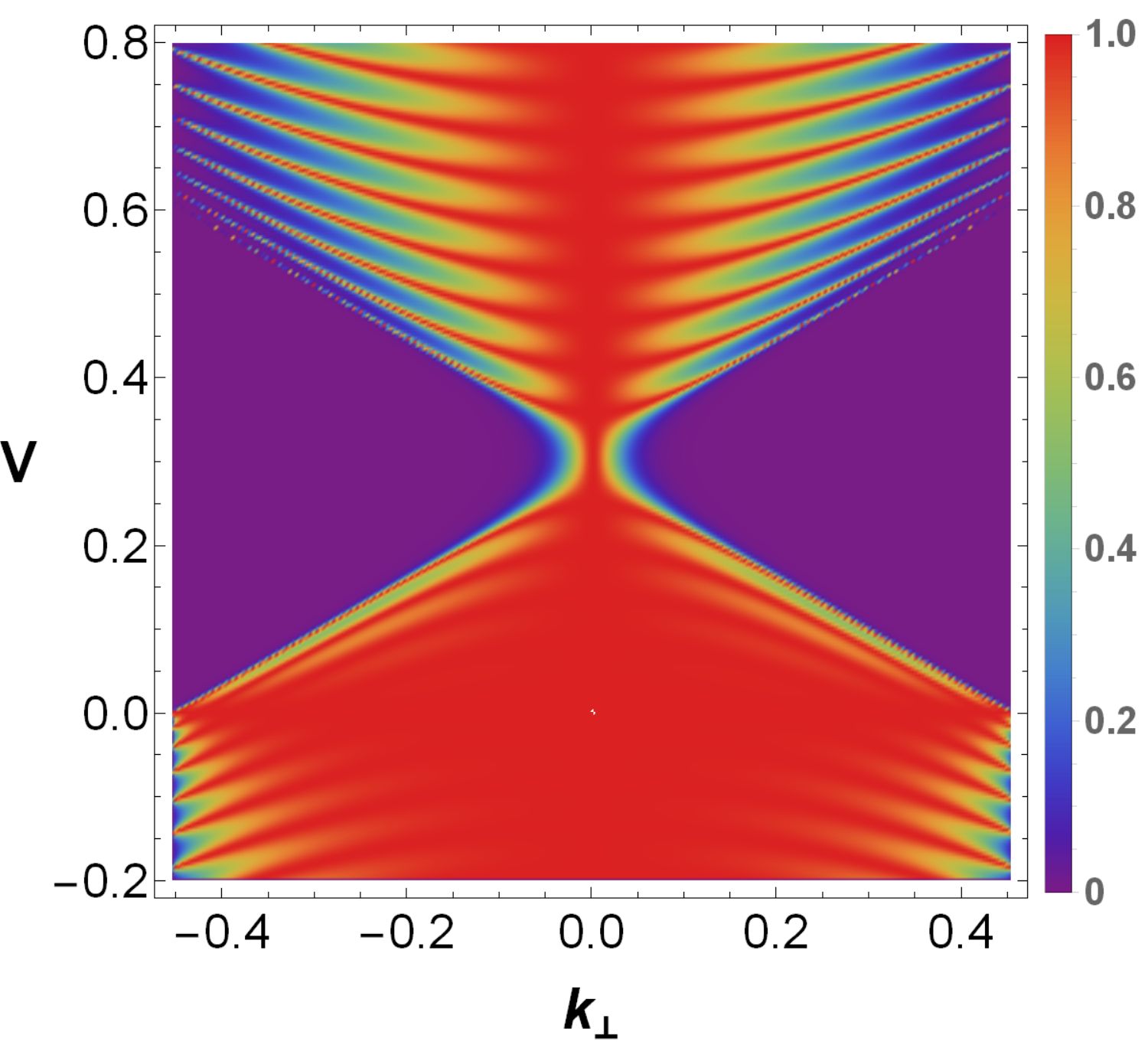}\label{Fig04c}}
	\subfigure[]{\includegraphics[height=3.5cm]{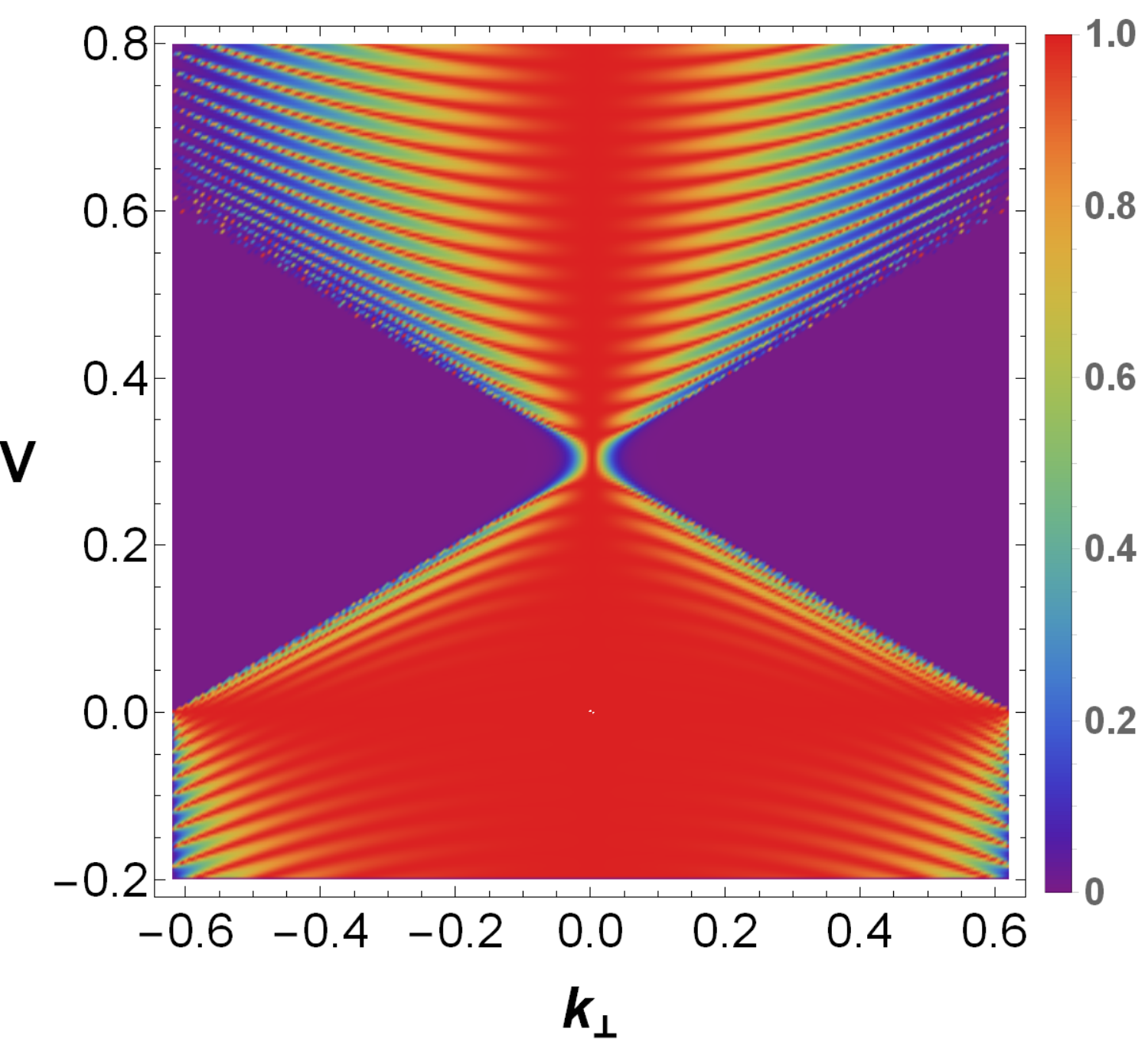}\label{Fig04d}}
	\caption{The barrier height dependence of the transmission probability in the $x$-direction for (a) $u=0$, (b) $u=\frac{1}{2}$, (c) $u=\frac{3}{4}$, and (d) $u=\frac{7}{8}$, with parameters $\varepsilon=0.3$, $\gamma=\pi/4$ and $d=8\pi$, and $k_{\perp}$ ranging from $-\frac{\varepsilon}{\sqrt{1-u^2}}$ to $\frac{\varepsilon}{\sqrt{1-u^2}}$.}
	\label{Fig04}
\end{figure}

\begin{figure}
	\centering
	\subfigure[]{\includegraphics[height=3.5cm]{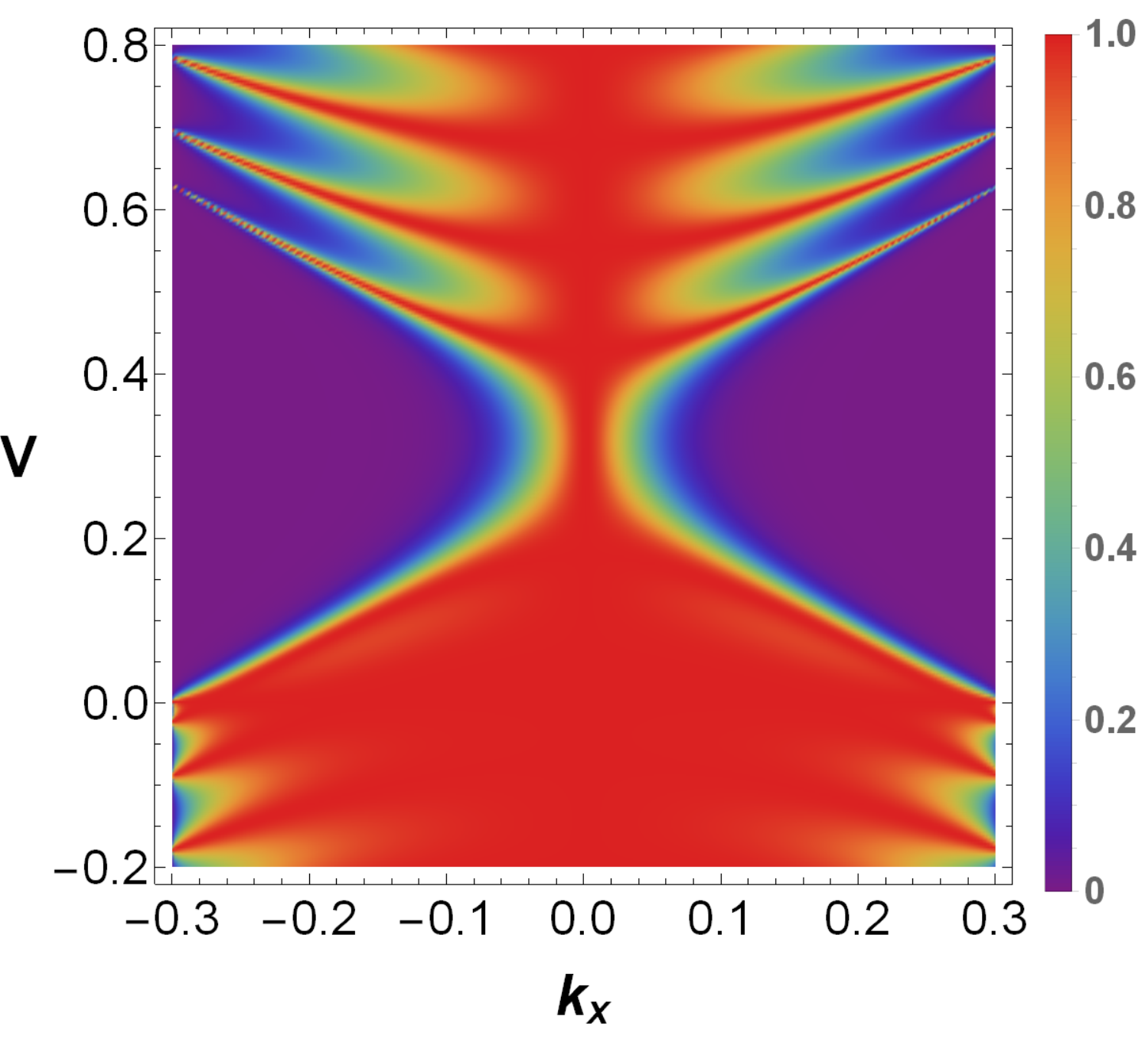}\label{Fig05a}}
	\subfigure[]{\includegraphics[height=3.5cm]{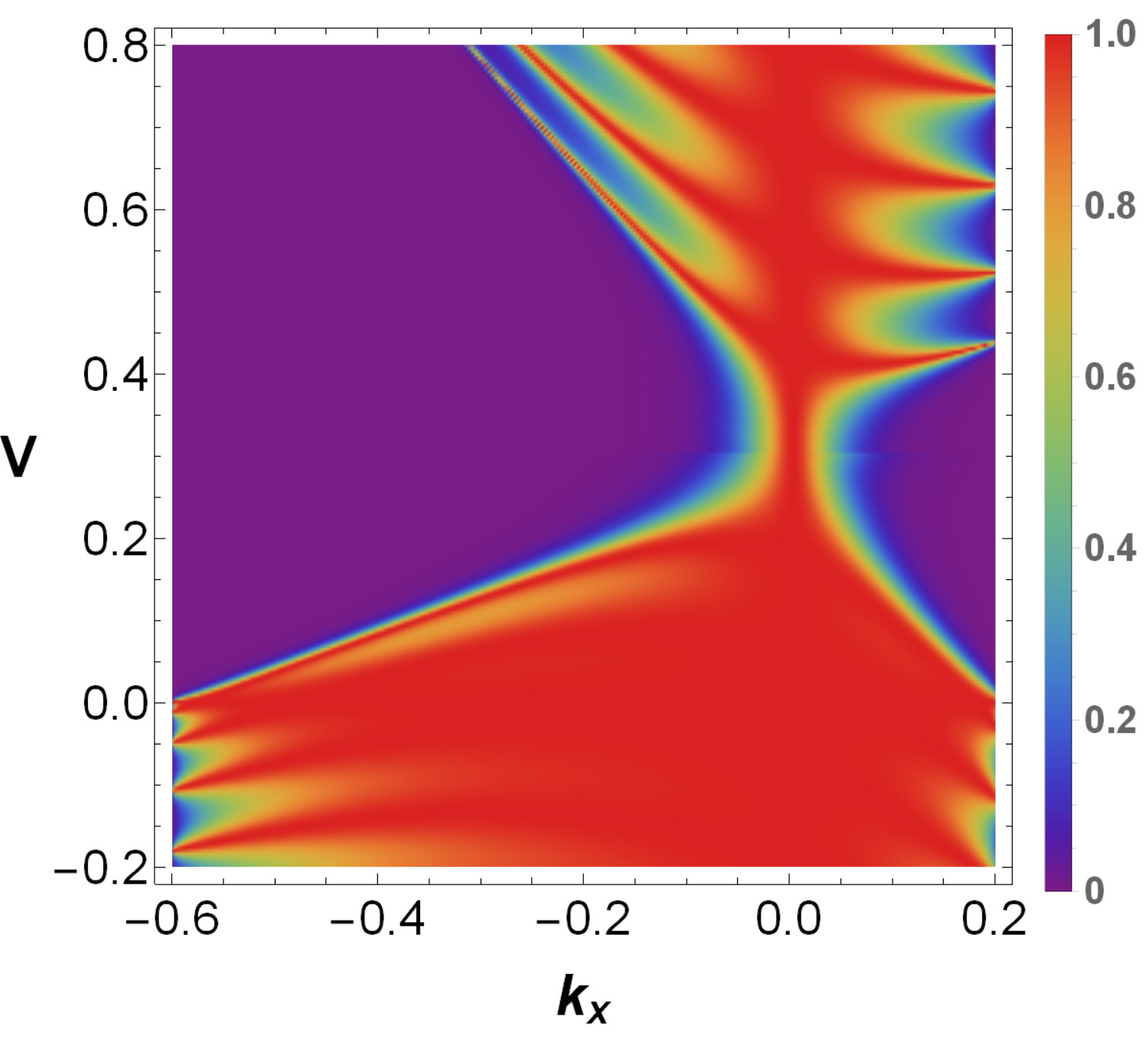}\label{Fig05b}}
	\subfigure[]{\includegraphics[height=3.5cm]{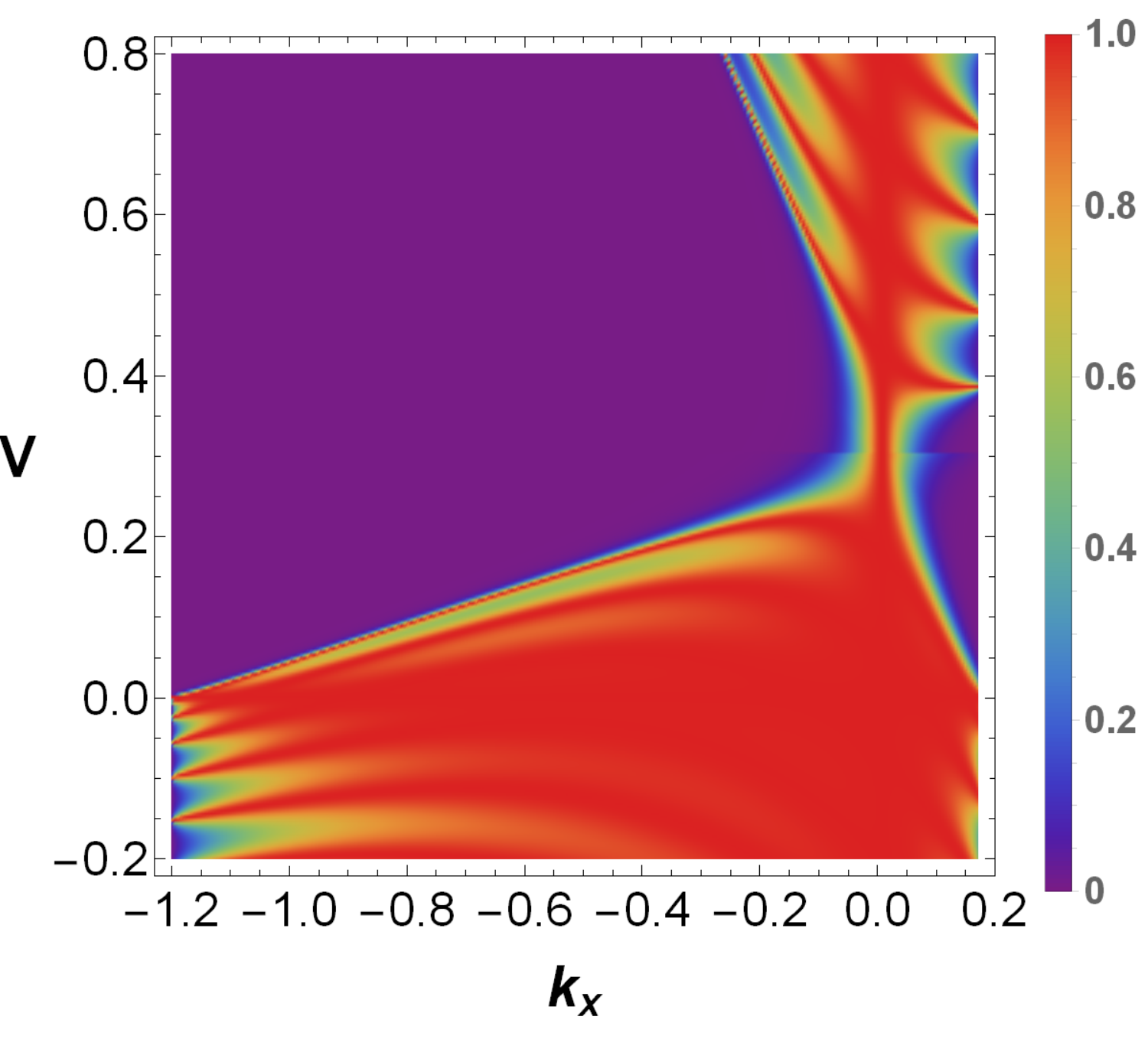}\label{Fig05c}}
	\subfigure[]{\includegraphics[height=3.5cm]{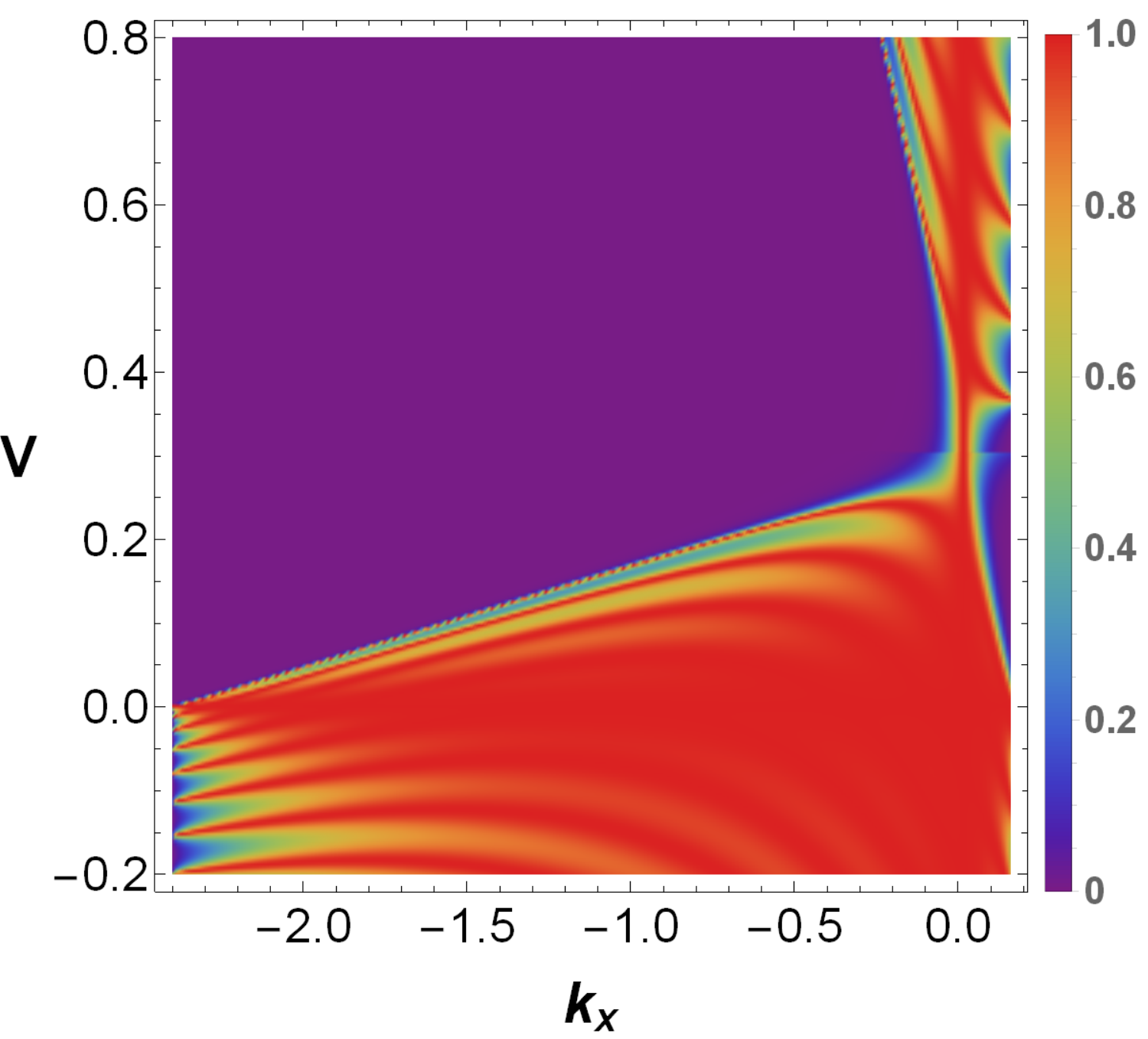}\label{Fig05d}}
	\caption{The barrier height dependence of the transmission probability in the $z$-direction for (a) $u=0$, (b) $u=\frac{1}{2}$, (c) $u=\frac{3}{4}$, and (d) $u=\frac{7}{8}$, with parameters $\varepsilon=0.3$, $\gamma=0$ and $d=8\pi$, and $k_{x}$ ranging from $\frac{\varepsilon}{-1+u}$ to $\frac{\varepsilon}{1+u}$.}
	\label{Fig05}
\end{figure}

\begin{figure}
	\centering
	\subfigure[]{\includegraphics[height=3.5cm]{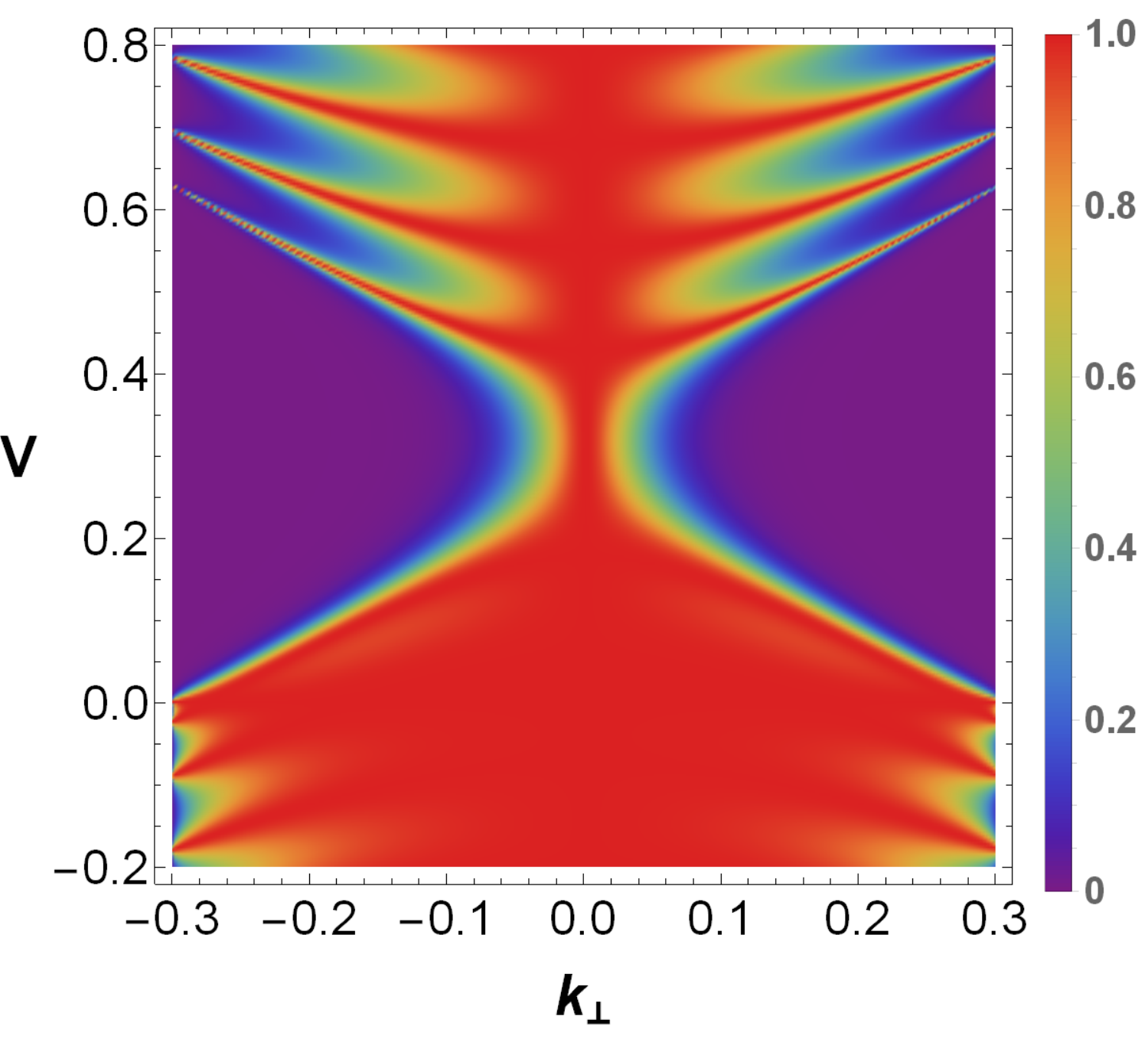}\label{Fig06a}}~~
	\subfigure[]{\includegraphics[height=3.5cm]{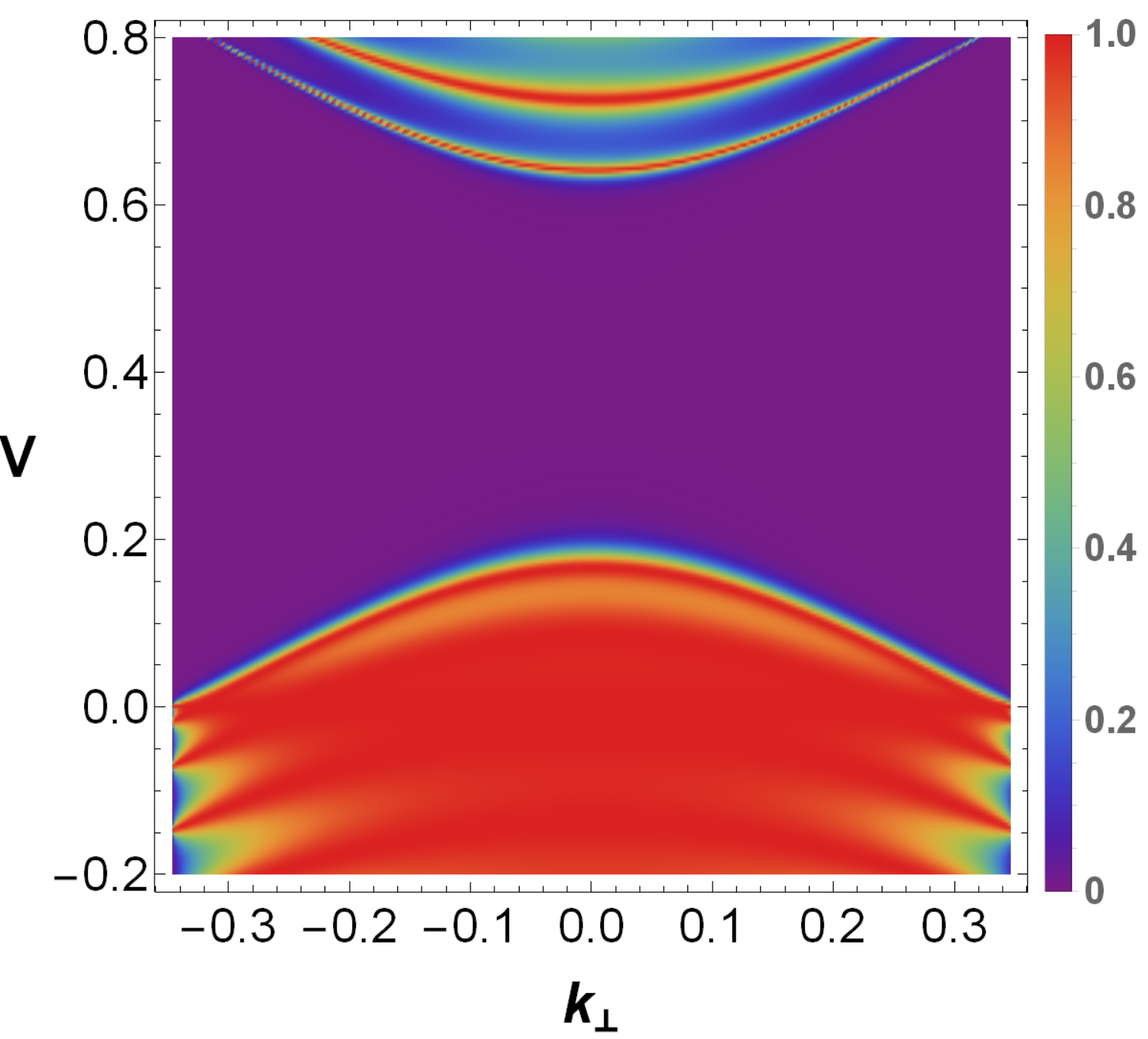}\label{Fig06b}}
	\subfigure[]{\includegraphics[height=3.5cm]{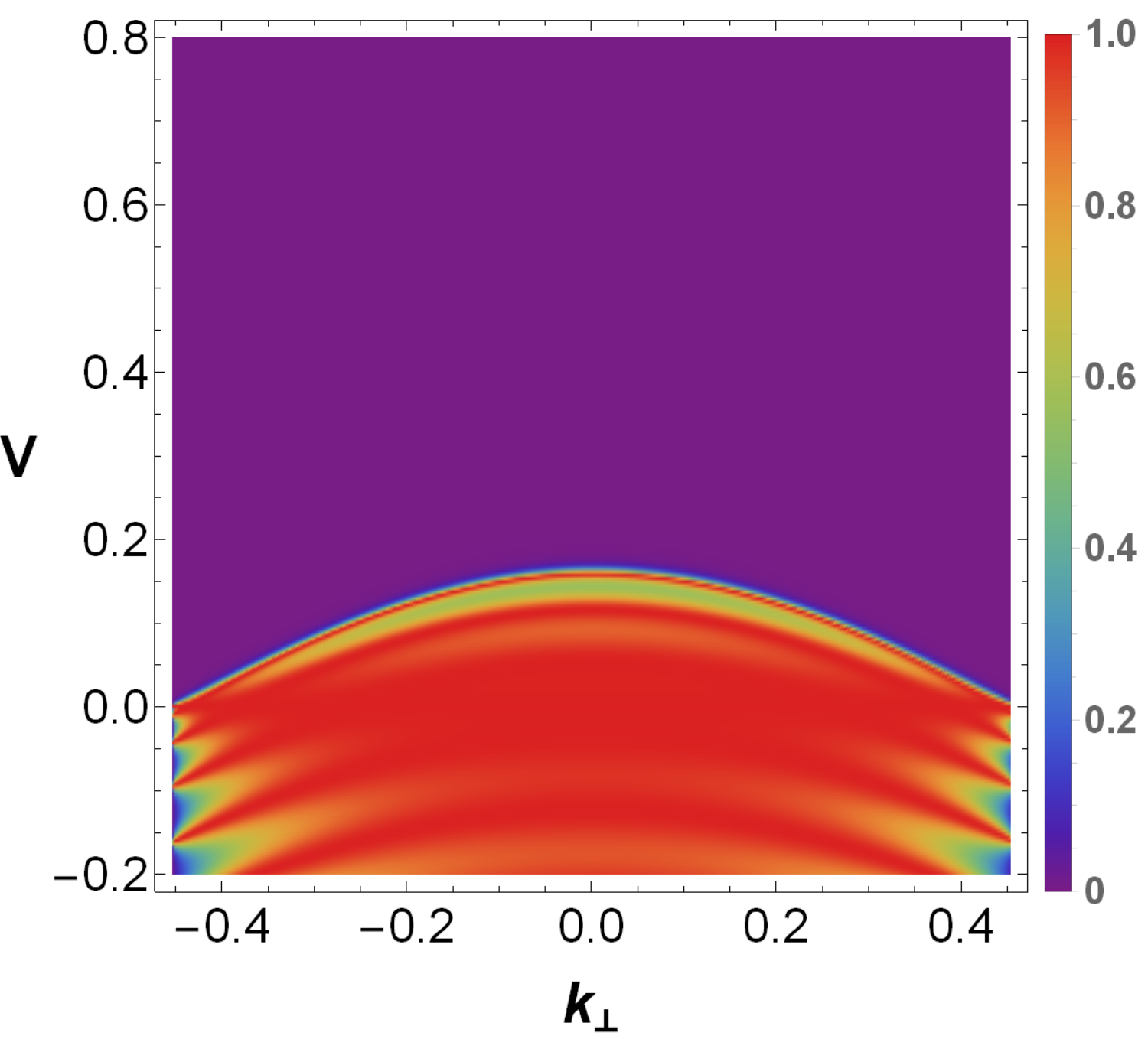}\label{Fig06c}}~~
	\subfigure[]{\includegraphics[height=3.5cm]{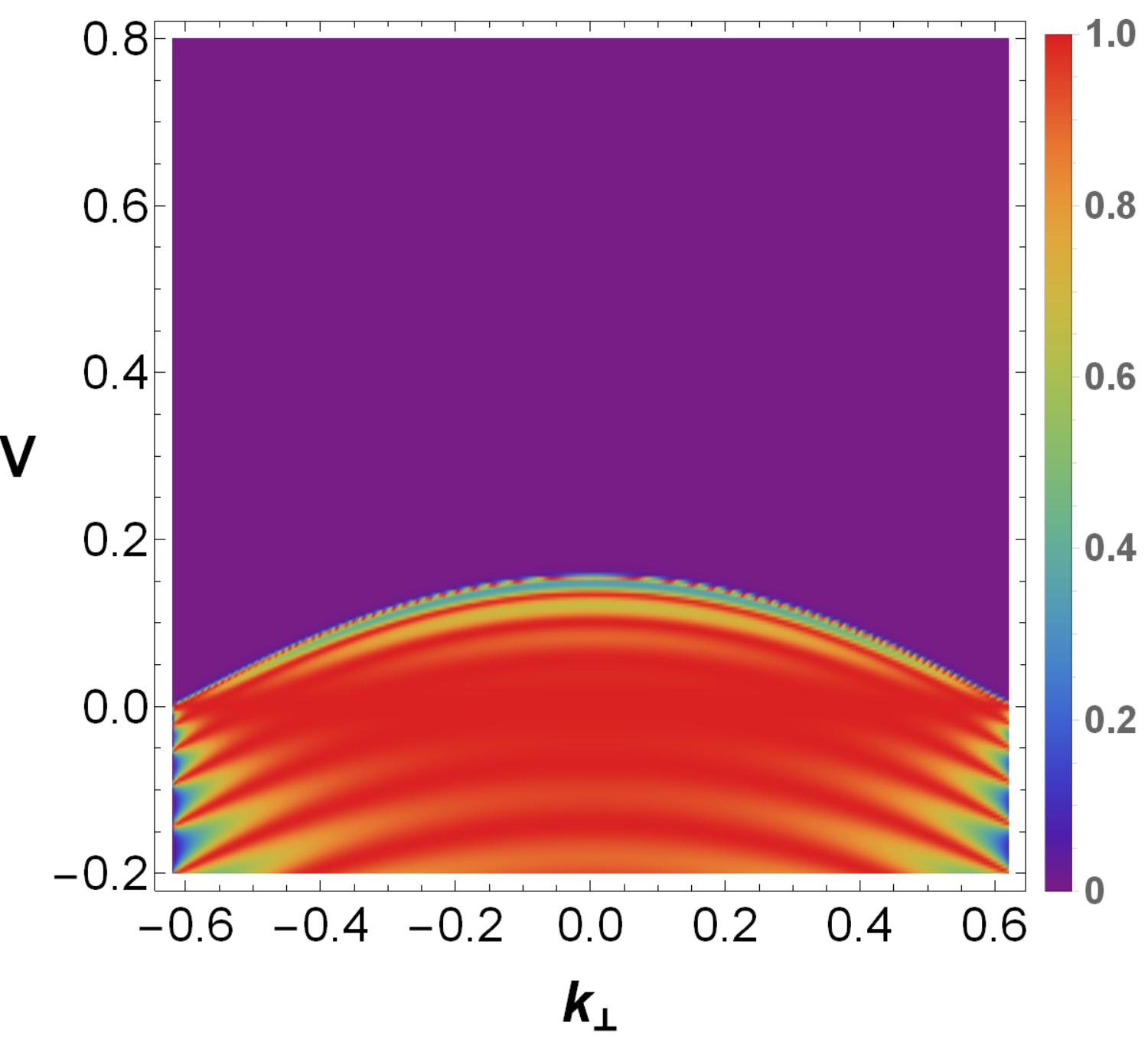}\label{Fig06d}}
	\caption{The barrier height dependence of the transmission probability in the $z$-direction for (a) $u=0$, (b) $u=\frac{1}{2}$, (c) $u=\frac{3}{4}$, and (d) $u=\frac{7}{8}$, with parameters $\varepsilon=0.3$, $\gamma=\pi/2$ and $d=8\pi$, and $k_{\perp}$ ranging from $-\frac{\varepsilon}{\sqrt{1-u^2}}$ to $\frac{\varepsilon}{\sqrt{1-u^2}}$.}
	\label{Fig06}
\end{figure}

When the tunneling happens in the $z$-direction, the wave vectors in the barrier region become
\begin{equation}
	\begin{aligned}
		q_{z,s}=s\sqrt{(\varepsilon-uk_x-V)^2-k_x^2-k_y^2}.
	\end{aligned}
\end{equation}
In this case, the incident magnons satisfy
\begin{equation}
	\begin{aligned}
		\frac{\varepsilon^2}{1-u^2}>(1-u^2)(k_x+\frac{\varepsilon u}{1-u^2})^2+k_y^2.
	\end{aligned}
	\label{ellispse}
\end{equation}
We introduce $k_{\perp}^2=(1-u^2)(k_x+\frac{\varepsilon u}{1-u^2})^2+k_y^2$ and $\gamma=\arcsin(k_y/k_{\perp})$, so
\begin{equation}
	\begin{aligned}
		k_{x} & =\frac{k_{\perp}\cos\gamma}{\sqrt{1-u^2}}-\frac{\varepsilon u}{1-u^2}, \\
		k_{y} & =k_{\perp}\sin\gamma.
	\end{aligned}
\end{equation}
The transmission probability is plotted for $\gamma=0$ and $\gamma=\pi/2$ in Fig.\ref{Fig05} and Fig.\ref{Fig06}, respectively. Note that $k_x=k_y=0$ is different from the normal incidence in real space, which is given by $dE/dk_x=dE/dk_y=0$. In fact, the normal incidence in real space corresponds to $k_x=-\frac{\varepsilon u}{1-u^2},k_y=0$, at which the transmission is not always perfect with different $V$\cite{Nguyen2018Klein}. However, at $k_x=k_y=0$ which is a focus of the elliptical boundary of the inequality (\ref{ellispse}), Klein tunneling with $T=1$ still occurs. We will refer to $k_x=k_y=0$ as the normal incidence in momentum space.

If the Weyl cone is tilted in the opposite direction ($-1<u<0$), Klein tunneling will be found at the other focus of the elliptical boundary of inequality (\ref{ellispse}). Actually, the symmetry in Klein tunneling $T(k_x,k_y,k_z)=T(-k_x,-k_y,-k_z)$ becomes $T(k_x,k_y,k_z)=T(k'_x,-k_y,-k_z)$, where $k'_x$ is given by Eq.(\ref{k'_x}).

\subsection{Double Weyl magnons}\label{SubSec:DWMagnon}
\begin{figure}[]
	\centering
	\includegraphics[height=4cm]{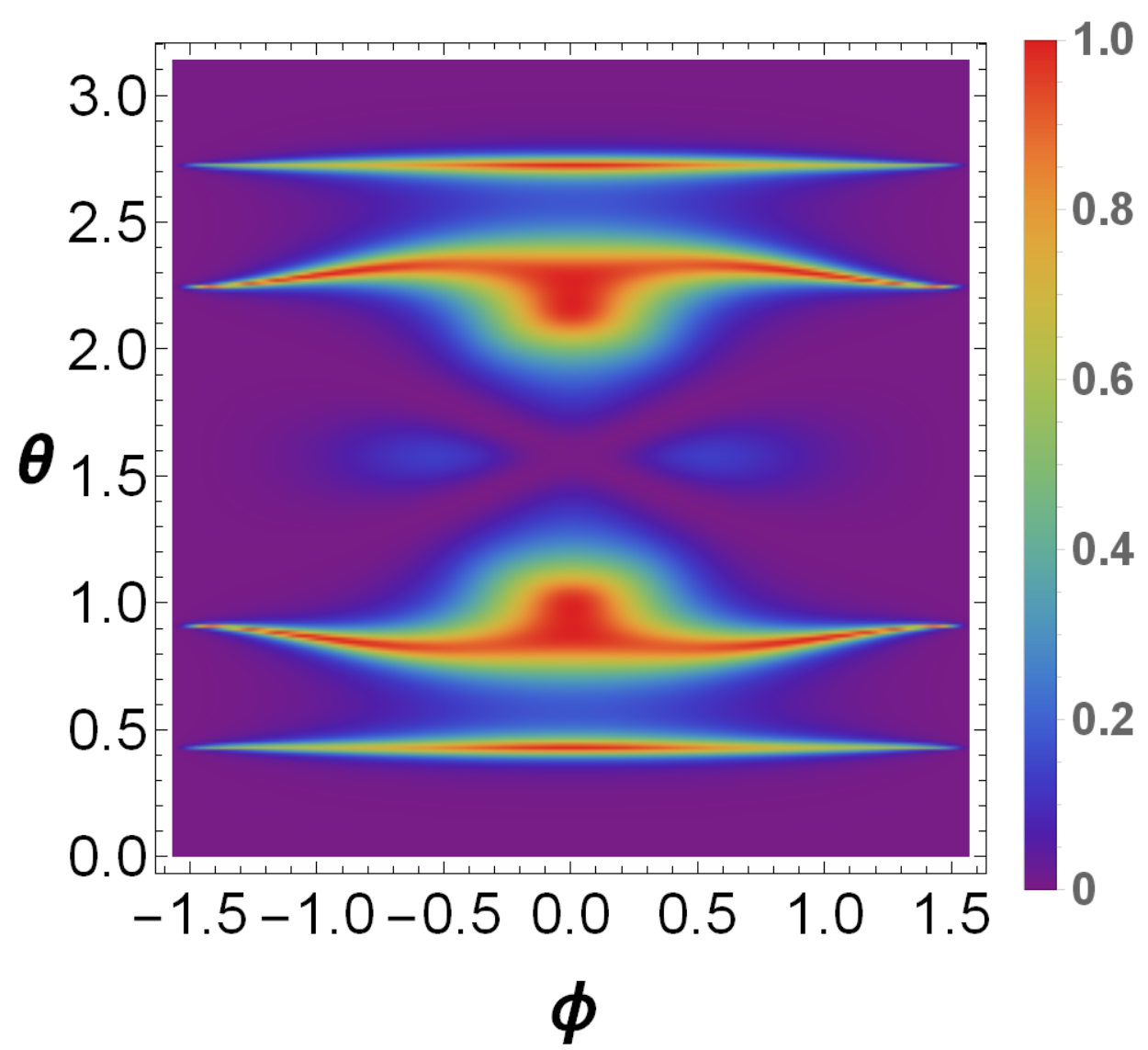}\label{Fig07a}
	\caption{The momentum angle dependence of the transmission probability of double Weyl magnons, in which $\varepsilon=0.3$, $V=0.6875$, and $d=8\pi$.}
	\label{Fig07}
\end{figure}

\begin{figure}[]
	\centering
    \subfigure[]{\includegraphics[height=3.5cm]{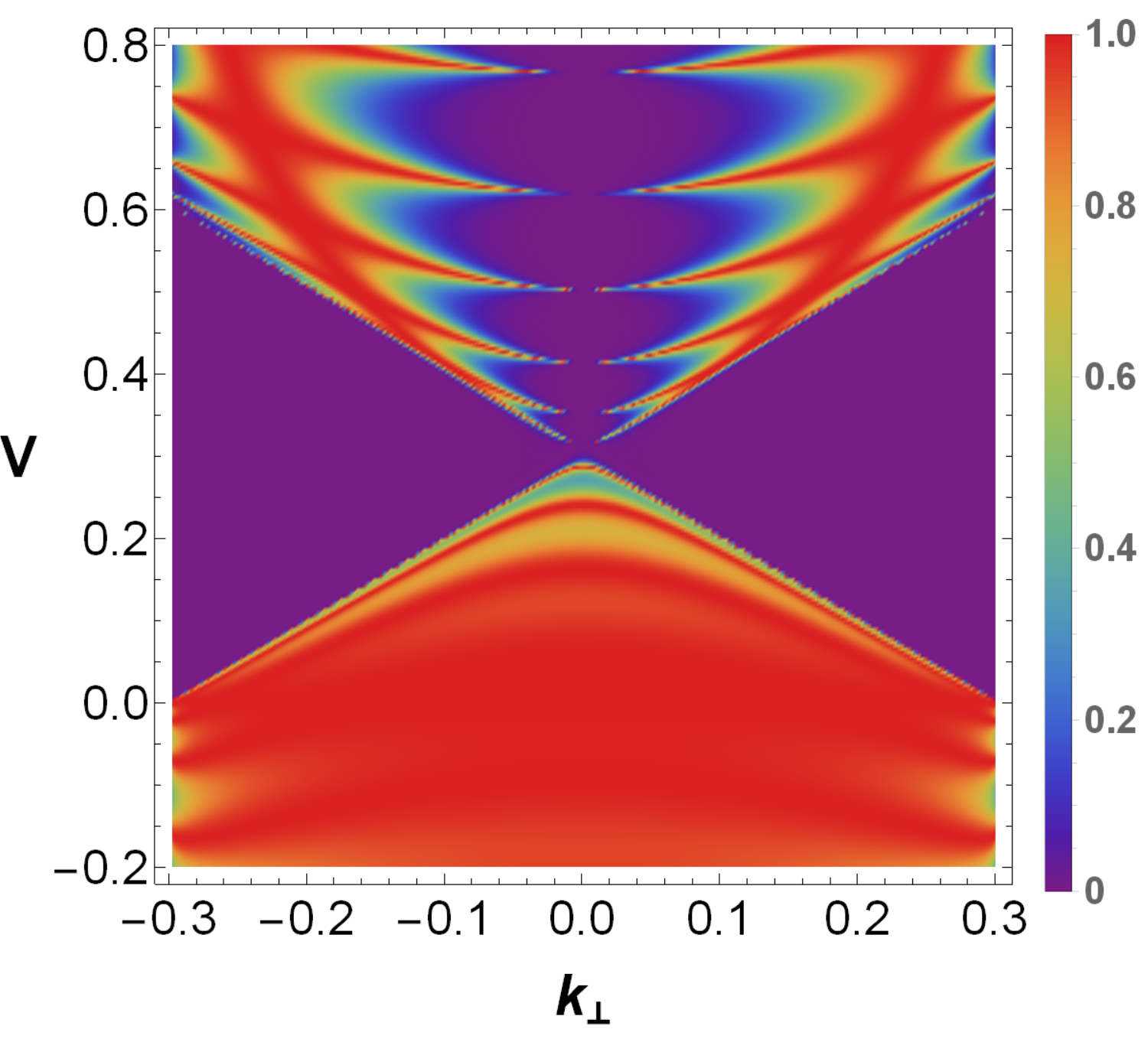}\label{Fig08a}}
	\subfigure[]{\includegraphics[height=3.5cm]{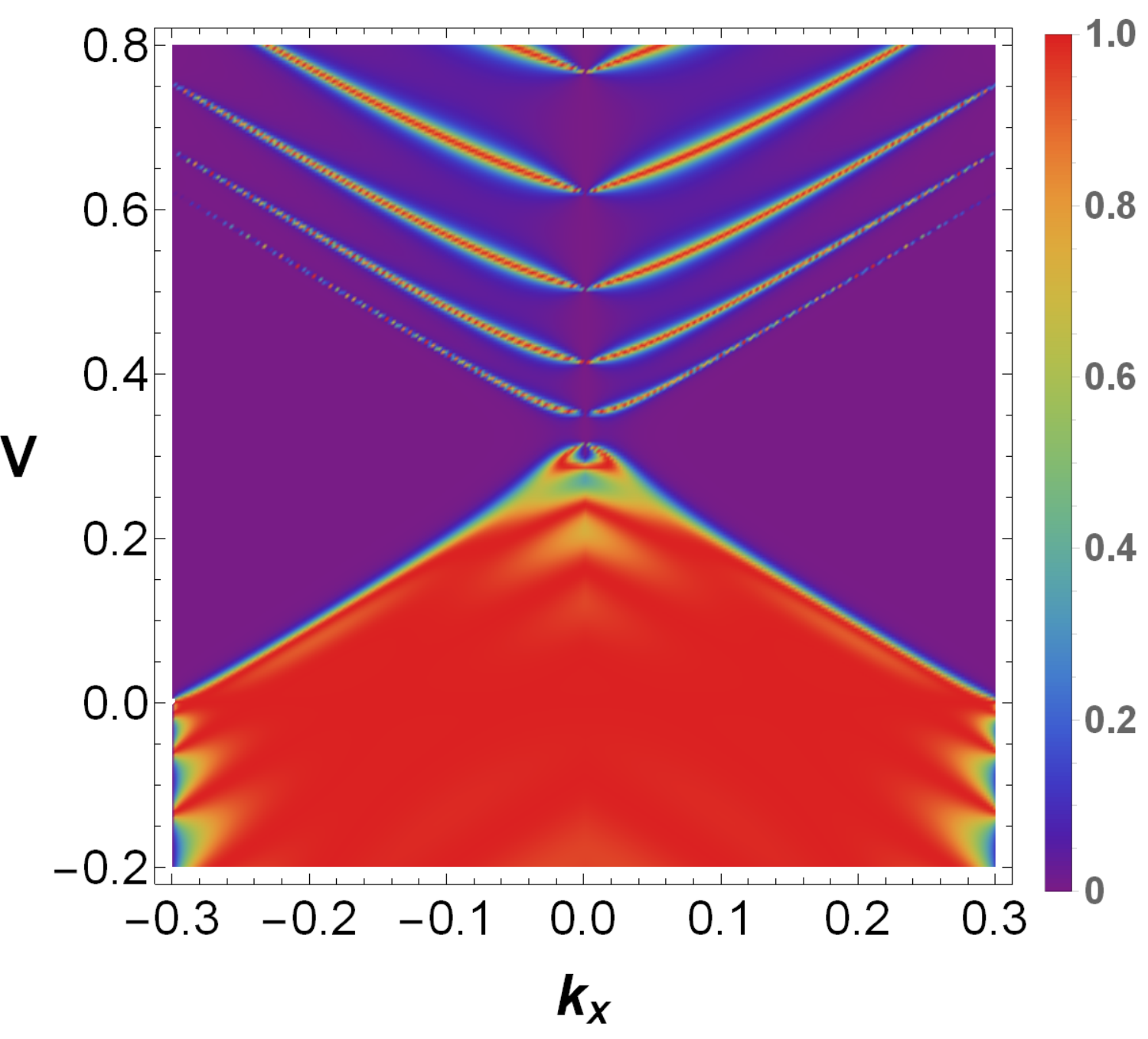}\label{Fig08b}}
	\caption{The barrier height dependence of the $x$-direction transmission probability of double Weyl magnons for (a) $\gamma=0$ and (b) $\gamma=\pi$. The parameters are $\varepsilon=0.3$ and $d=8\pi$, and $k_{\perp}$ ranges from $-\varepsilon$ to $\varepsilon$.}
	\label{Fig08}
\end{figure}

Another interesting derivative of Weyl materials is the multi-Weyl materials, in which the multi-Weyl points can be viewed as the merger of Weyl points with the same chirality\cite{Fang2012PRL}. Again, we first consider the tunneling happening in the $x$-direction. In this case, the effective Hamiltonian can be expressed as
\begin{equation}
	\begin{aligned}
		H_{MW}({\bf k}) & =k^J_{+}\sigma_{-}+k^J_{-}\sigma_{+}+k_{z}\sigma_{z}+V(x)\sigma_{0} \\
		          & =\left(
		\begin{array}{cc}
				k_z+V(x)      & (k_x-i k_y)^J \\
				(k_x+i k_y)^J & -k_z+V(x)     \\
			\end{array}
		\right),
	\end{aligned}
	\label{DW}
\end{equation}
in which $k_{\pm}=k_x\pm i k_y$, $\sigma_{\pm}=\frac{1}{2}(\sigma_x\pm i \sigma_y)$, and $J=1,2,3$. The eigenenergy is
\begin{equation}
	\begin{aligned}
		E_{MW}({\bf k})=s\sqrt{(k_x^2+k_y^2)^J+k_z^2}+V(x),
	\end{aligned}
\end{equation}
where $s=\pm1$. The wave function of multi-Weyl magnons with energy $\varepsilon$ is
\begin{equation}
	\begin{aligned}
		\psi(x) & =
            \frac{1}{\sqrt{A'}}\left[
			\begin{array}{c}
				k_z+\varepsilon \\
				(k_x+i k_y)^{J}
			\end{array}
			\right]e^{ik_xx} \equiv\left[
			\begin{array}{c}
				\psi_{1}(k_x) \\
				\psi_{2}(k_x)
			\end{array}
			\right]e^{ik_xx},
	\end{aligned}
	\label{wfmw}
\end{equation}
where $A'=2\varepsilon(\varepsilon+k_z)$. Therefore, the wave functions in the three regions are, respectively,
\begin{equation}
	\begin{aligned}
		\psi^{\mathrm{i}}(x) & =\left[
			\begin{array}{c}
				\psi_{1}(k_x) \\
				\psi_{2}(k_x)
			\end{array}
			\right]e^{ik_xx}+r\left[
			\begin{array}{c}
				\psi_{1}(-k_x) \\
				\psi_{2}(-k_x)
			\end{array}
			\right]e^{-ik_xx}                                 \\&+\sum_{n=1}^{J-1}r_n\left[
			\begin{array}{c}
				\psi_{1}(-K_{x}^{(n)}) \\
				\psi_{2}(-K_{x}^{(n)})
			\end{array}
			\right]e^{-iK_{x}^{(n)}x},                            \\
		\psi^{\mathrm{p}}(x) & =\sum_{n=0}^{J-1}a_n\left[
			\begin{array}{c}
				\psi_{1}(Q_{x}^{(n)}) \\
				\psi_{2}(Q_{x}^{(n)})
			\end{array}
			\right]e^{iQ_{x}^{(n)}x}                              \\&+\sum_{n=0}^{J-1}b_n\left[
			\begin{array}{c}
				\psi_{1}(-Q_{x}^{(n)}) \\
				\psi_{2}(-Q_{x}^{(n)})
			\end{array}
			\right]e^{-iQ_{x}^{(n)}x},                            \\
		\psi^{\mathrm{t}}(x) & =t\left[
			\begin{array}{c}
				\psi_{1}(k_x) \\
				\psi_{2}(k_x)
			\end{array}
			\right]e^{ik_xx}+\sum_{n=1}^{J-1}t_n\left[
			\begin{array}{c}
				\psi_{1}(K_{x}^{(n)}) \\
				\psi_{2}(K_{x}^{(n)})
			\end{array}
			\right]e^{iK_{x}^{(n)}x}.
	\end{aligned}
	\label{WaveMWX}
\end{equation}
where
\begin{equation}
	\begin{aligned}
	K_{x}^{(n)}&=\sqrt{\left(\varepsilon^2-k_z^2\right)^\frac{1}{J}e^{i\frac{2\pi n}{J}}-k_y^2},\\
    Q_{x}^{(n)}&=\sqrt{\left[(\varepsilon-V)^2-k_z^2\right]^\frac{1}{J}e^{i\frac{2\pi n}{J}}-k_y^2}.
    \end{aligned}
	\label{kxmw}
\end{equation}
Note that $K_x^{(n)}$ has a positive imaginary part, and we have chosen the associated waves to be evanescent waves which decay exponentially away from the potential barrier. There are $4J$ coefficients to be solved for, so $2J$ boundary conditions at each of the two boundaries are required. Here, we focus on the double Weyl case, i.e. $J=2$. By integrating the eigenequation over an infinitesimal region across the boundary ($x=0$ or $x=d$), we see that the derivative of the wave functions must be continuous. Therefore, we have
\begin{equation}
	\begin{aligned}
		\psi^{\mathrm{i}}(0)             & =\psi^{\mathrm{p}}(0),             \\
		\frac{d\psi^{\mathrm{i}}(x)}{dx}\Bigg|_{x=0} & =\frac{d\psi^{\mathrm{p}}(x)}{dx}\Bigg|_{x=0}, \\
		\psi^{\mathrm{p}}(d)             & =\psi^{\mathrm{t}}(d),             \\
		\frac{d\psi^{\mathrm{p}}(x)}{dx}\Bigg|_{x=d} & =\frac{d\psi^{\mathrm{t}}(x)}{dx}\Bigg|_{x=d},
	\end{aligned}
	\label{conditions}
\end{equation}
which are actually eight equations. Therefore, the transmission probability $T=|t|^2$ can be found. Fig.\ref{Fig07} shows the momentum angle dependence of $T$ with the spherical coordinates given in Eq.(\ref{SphericalCoordinates}). Meanwhile, all possible incident magnons satisfy
\begin{equation}
	\begin{aligned}
		\varepsilon^2>k_y^4+k_z^2.
	\end{aligned}
	\label{ConditionKDWX}
\end{equation}
We introduce $k_{\perp}^2=k_y^4+k_z^2$ and $\gamma=\arcsin(k_z/k_{\perp})$. The transmission probability is plotted for $\gamma=0$ and $\gamma=\pi$ in Fig.\ref{Fig08}. As shown in Figs.\ref{Fig07} and \ref{Fig08}, in the double Weyl cone case, the Klein tunneling effect in the $x$-direction disappears due to the emergence of the evanescent waves. However, the Fabry-P\'erot resonances still exist and cause interference fringes. In fact, when $k_z=0$, Eq.(\ref{DW}) comes back to the Schrödinger equation. And there is no barrier height independent tunneling for the normal incident in this case.


When the tunneling happens in the $z$-direction, the wave function is similar to Eq.(\ref{wfmw}),
\begin{equation}
	\begin{aligned}
		\psi(z)
		 & =\left[
			\begin{array}{c}
				\psi_{1}(k_z) \\
				\psi_{2}(k_z)
			\end{array}
			\right]e^{ik_zz}.
	\end{aligned}
	\label{wfmwz}
\end{equation}
The wave functions in the three regions are, respectively,
\begin{equation}
	\begin{aligned}
		\psi^{\mathrm{i}}(z) & =\left[
			\begin{array}{c}
				\psi_{1}(k_z) \\
				\psi_{2}(k_z)
			\end{array}
			\right]e^{ik_zz}+r\left[
			\begin{array}{c}
				\psi_{1}(-k_z) \\
				\psi_{2}(-k_z)
			\end{array}
			\right]e^{-ik_zz},              \\
		\psi^{\mathrm{p}}(z) & =a\left[
			\begin{array}{c}
				\psi_{1}(q_z) \\
				\psi_{2}(q_z)
			\end{array}
			\right]e^{iq_zz}+b\left[
			\begin{array}{c}
				\psi_{1}(-q_z) \\
				\psi_{2}(-q_z)
			\end{array}
			\right]e^{-iq_zz},              \\
		\psi^{\mathrm{t}}(z) & =t\left[
			\begin{array}{c}
				\psi_{1}(k_z) \\
				\psi_{2}(k_z)
			\end{array}
			\right]e^{ik_zz},
	\end{aligned}
	\label{WaveFZmw}
\end{equation}
where the wave vectors are
\begin{equation}
	\begin{aligned}
		k_z & =\sqrt{\varepsilon^2-(k_x^2+k_y^2)^J},     \\
		q_z & =\sqrt{(\varepsilon-V)^2-(k_x^2+k_y^2)^J},
	\end{aligned}
	\label{kzmw}
\end{equation}
given that the energy of the incident magnons is $\varepsilon$. Here, the continuity of wave functions Eq.(\ref{continuity}) is enough to solve for all the coefficients. Moreover, all possible incident magnons satisfy
\begin{equation}
	\begin{aligned}
		|\varepsilon|>k_x^2+k_y^2.
	\end{aligned}
	\label{ConditionKDWZ}
\end{equation}
We introduce $k_{\perp}^2=k_x^2+k_y^2$ and $\gamma=\arcsin(k_y/k_{\perp})$.

\begin{figure}[]
	\centering
    \subfigure[]{\includegraphics[height=3.5cm]{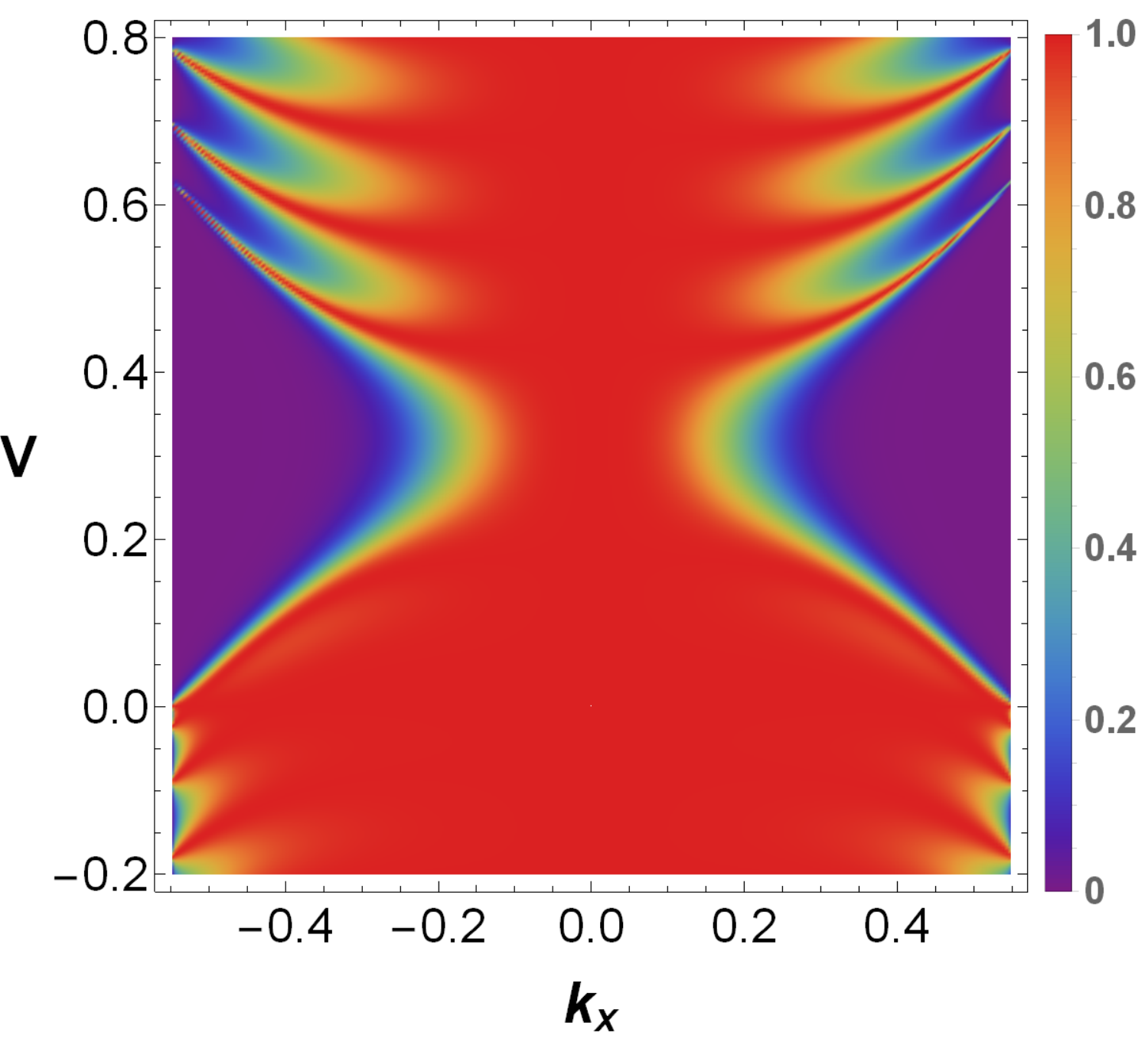}\label{Fig09a}}
	\subfigure[]{\includegraphics[height=3.5cm]{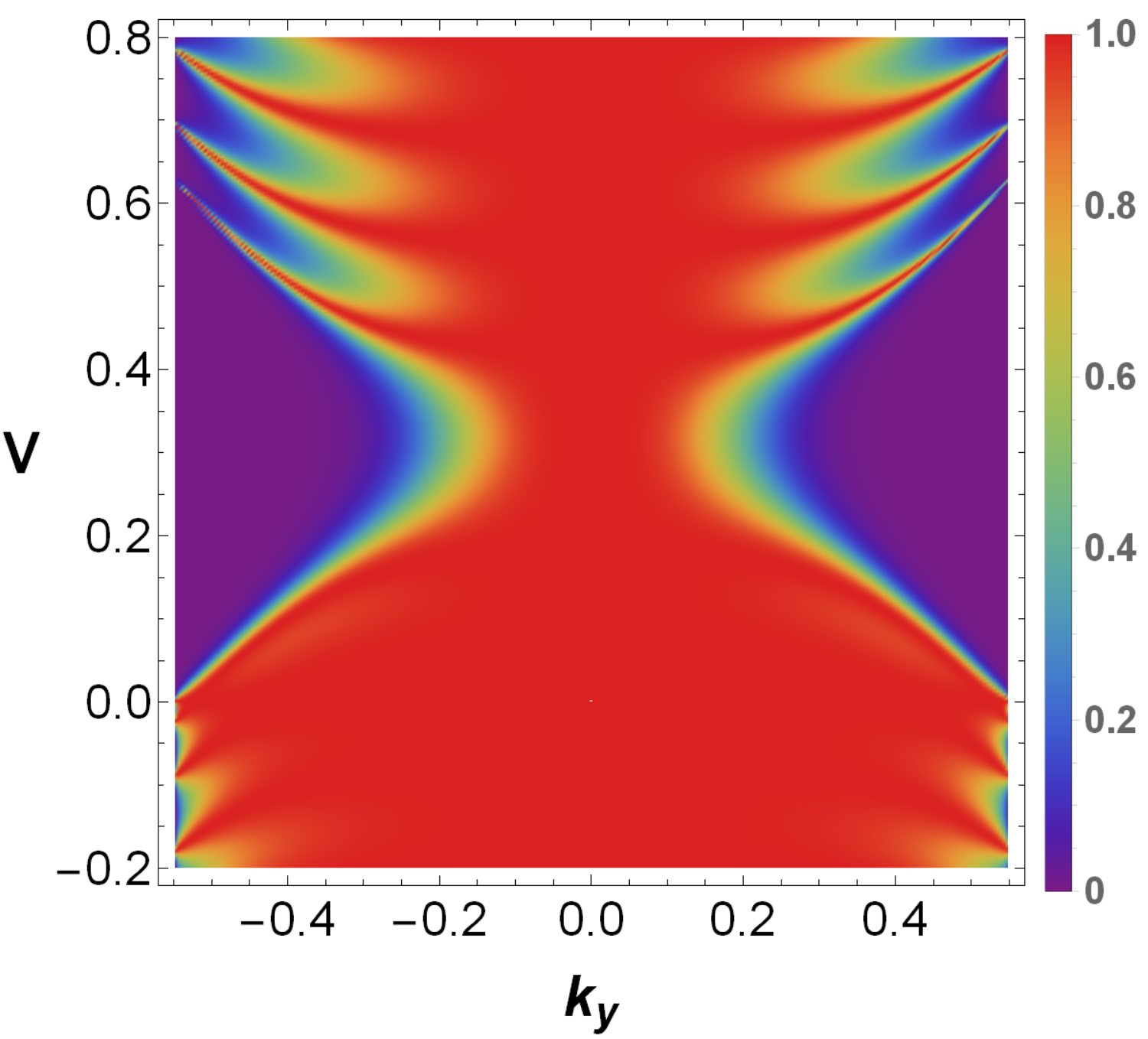}\label{Fig09b}}
	\caption{The barrier height dependence of the $z$-direction transmission probability of double Weyl magnons with (a) $\gamma=0$ and (b) $\gamma=\pi$. The parameters are $\varepsilon=0.3$ and $d=8\pi$, and $k_{\perp}$ ranges from $-\sqrt{\varepsilon}$ to $\sqrt{\varepsilon}$.}\label{Fig09}
\end{figure}

Moreover, with the absence of the evanescent waves, the Klein tunneling effect appears in the $z$-direction. Actually, the Klein tunneling effect is even stronger as there is a larger area with $V$-independent high transmission probability in Fig.\ref{Fig09}.

\section{Magnetization current carried by Weyl magnons}
\label{Sec:MConductance}
\begin{figure}[ht]
	\centering
	\subfigure[]{\includegraphics[height=3.cm]{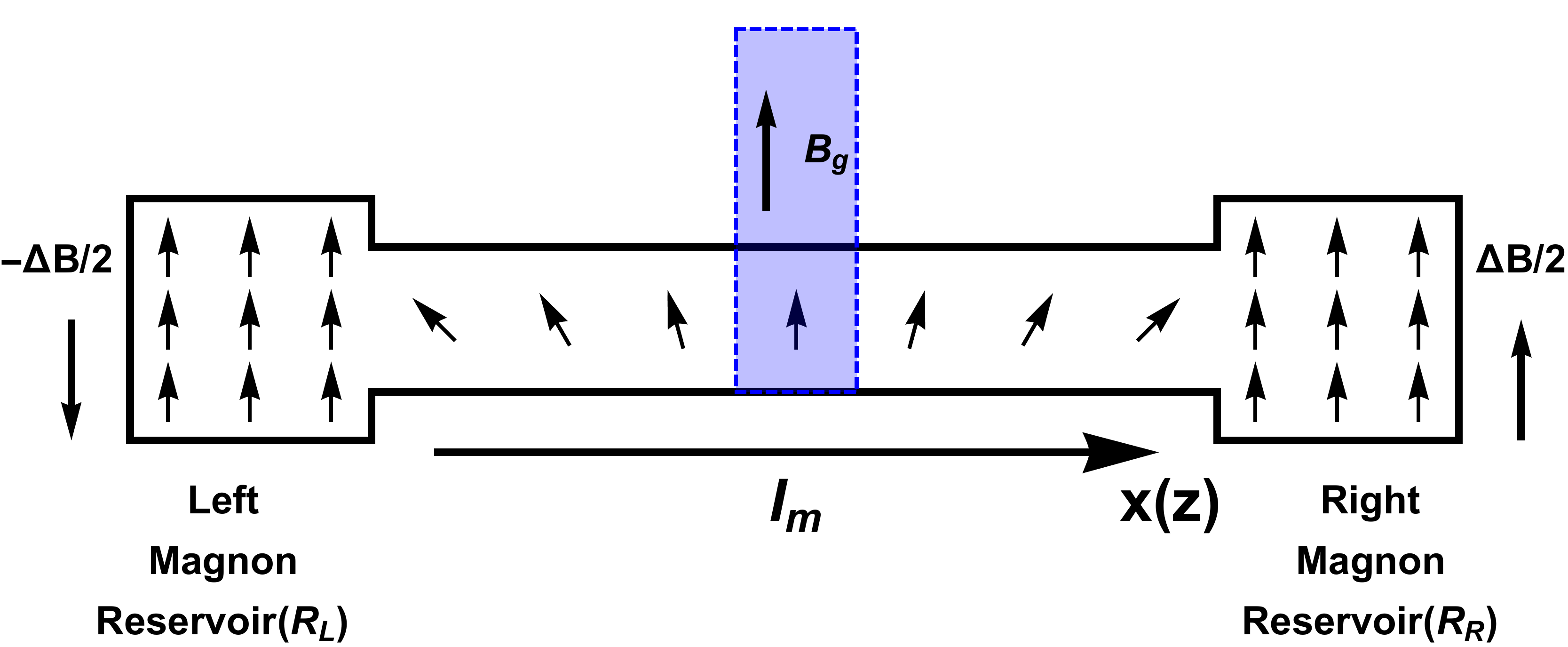}\label{Fig10a}}
	\subfigure[]{\includegraphics[height=2cm]{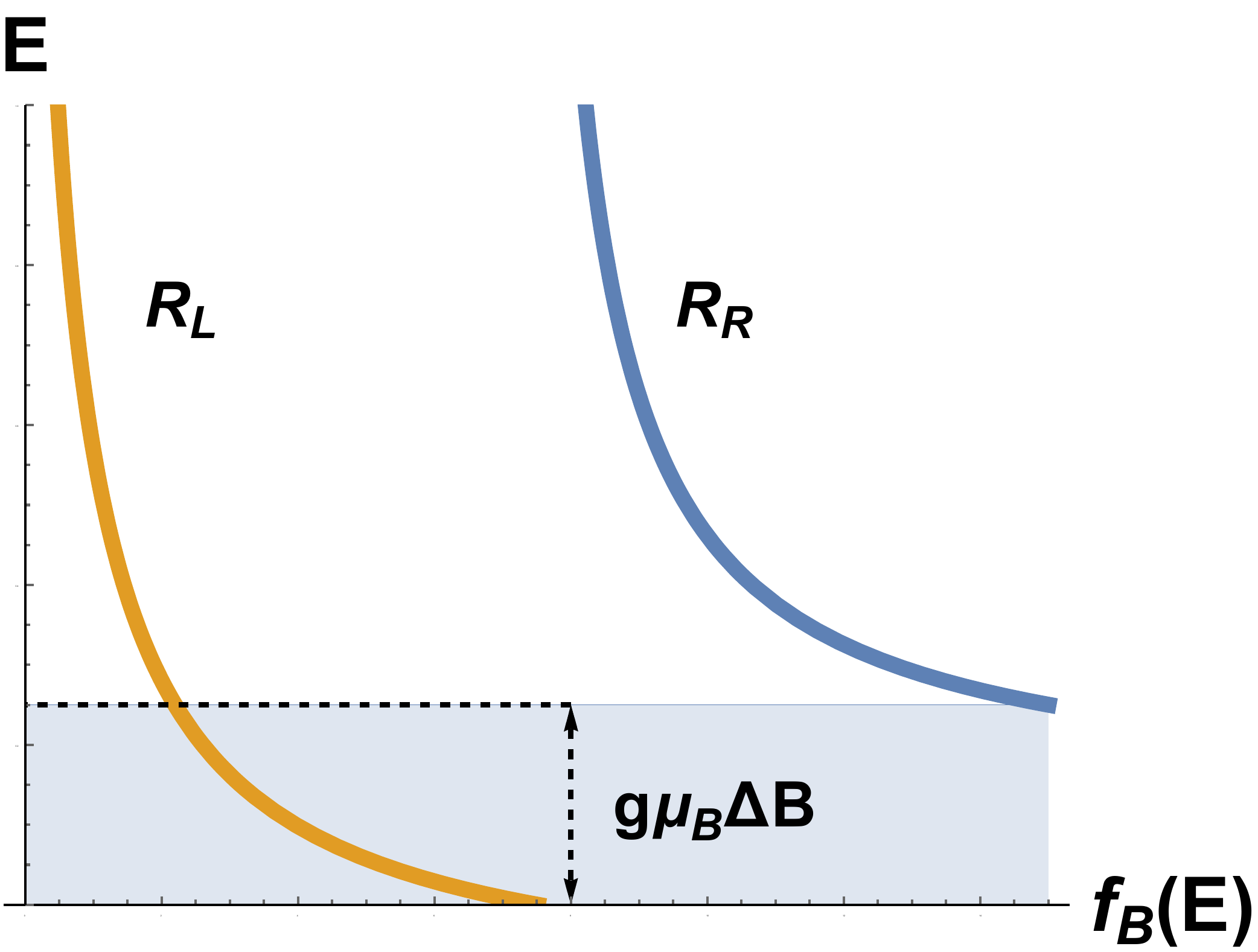}\label{Fig10b}}
	\subfigure[]{\includegraphics[height=2cm]{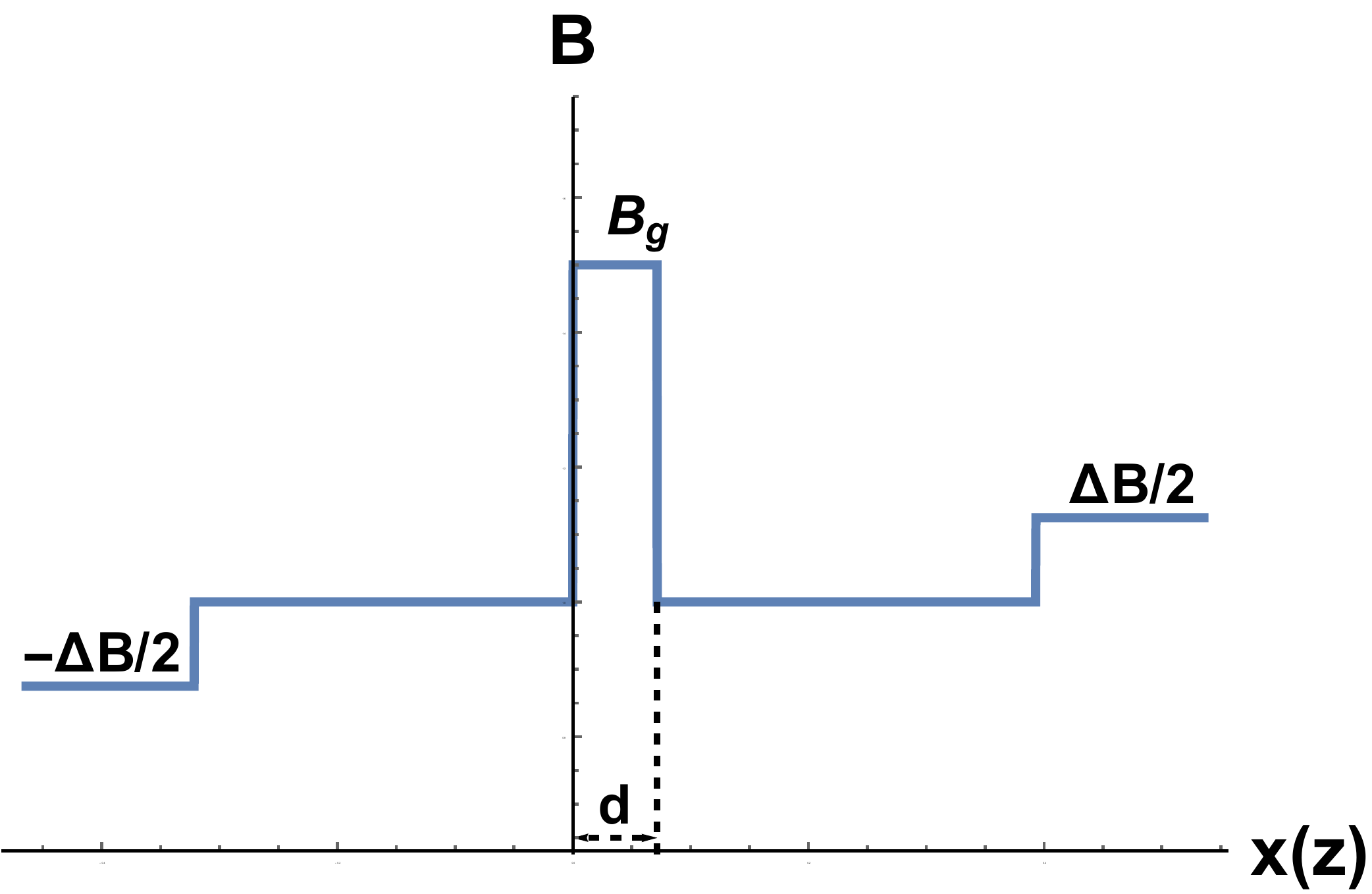}\label{Fig10c}}
	\subfigure[]{\includegraphics[height=2cm]{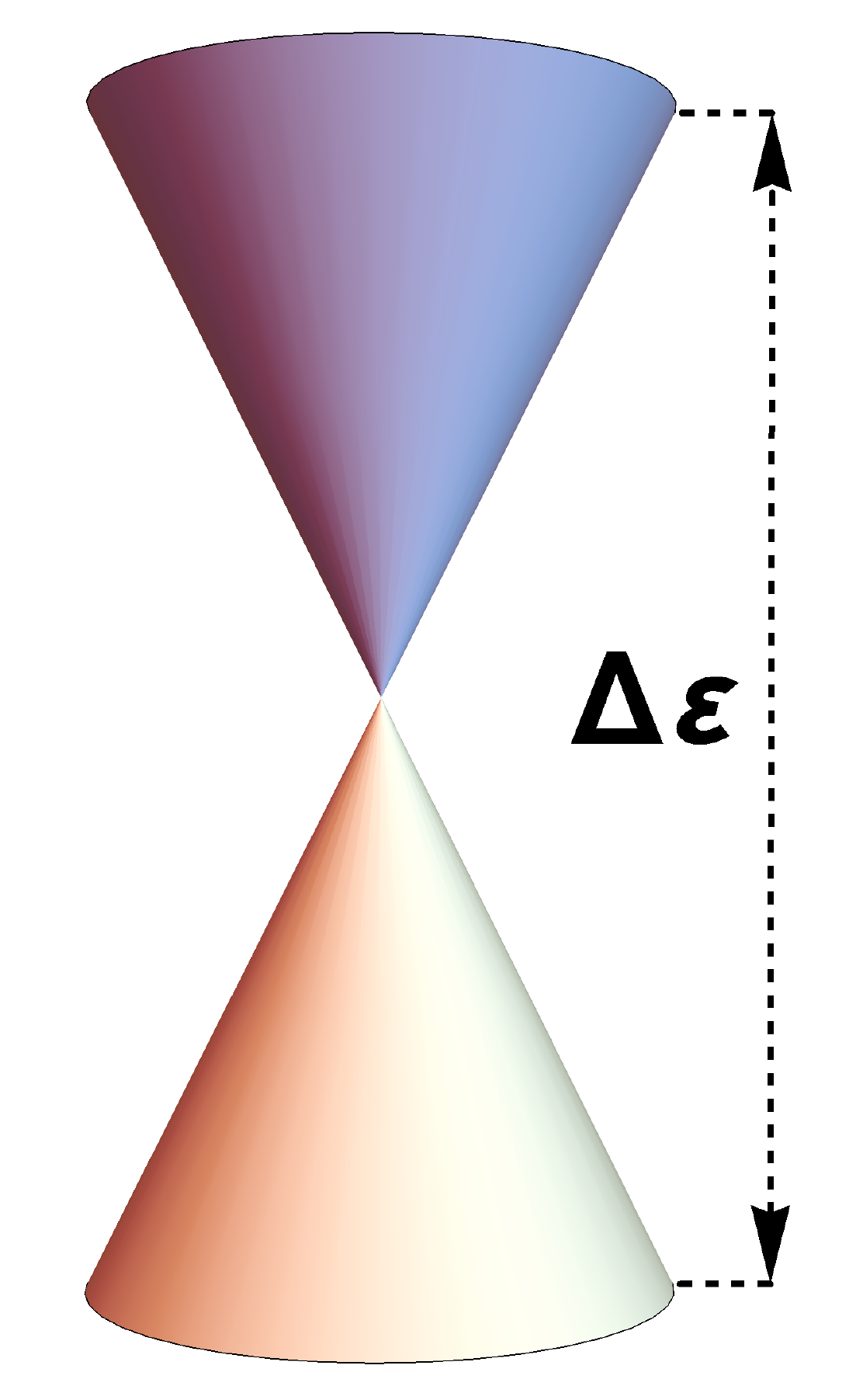}\label{Fig10d}}
	\caption{(a) Schematic diagram of the magnon transport in a magnetic wire in the $x$- or $z$-direction. The wire is made of magnonic Weyl materials\cite{Su2017MagnonicWeyl}. A gate magnetic field $B_{g}$ with width $d$ is applied to the wire, generating a potential barrier with height $V=\textsl{g}\mu_{B}B_{g}$ for the magnons. Driven by a magnetic field difference between the two bulk systems, a magnetization current $I_m$ carried by Weyl magnons goes from the left magnon reservoir $R_L$ to the right one $R_R$. (b) $\Delta B$ shifts the Bose distribution functions $f_{B}(E)$ in the reservoirs. Magnons with energies within the shaded region in $R_L$ are not transmitted to $R_R$\cite{Loss2003MagnetizationTransport}. (c) The variation of the external magnetic field along the magnetic wire. (d) Schematic diagram of the Weyl cone with energy range $\Delta\varepsilon$ in the magnon bands.}\label{Fig10}
\end{figure}

After giving a picture of the tunneling properties of Weyl magnons, we now consider their ballistic transport in a quasi-one-dimensional magnetic wire. The schematic diagram is given in Fig.\ref{Fig10}. The magnetic wire along the $x$- or $z$-direction as well as the reservoirs are made of magnonic Weyl materials\cite{Su2017MagnonicWeyl}. Driven by a magnon chemical potential difference between the two reservoirs, a magnetization current $I_m$ carried by magnons with a magnetic moment $-\textsl{g}\mu_{B}\textbf{e}_z$ goes from the left reservoir $R_L$ to the right one $R_R$\cite{Loss2003MagnetizationTransport}. Moreover, by applying a constant gate magnetic field $B_g$, a potential barrier with height $V=\textsl{g}\mu_{B}B_{g}$ and width $d$ is generated, shown as the blue region in Fig.\ref{Fig10a}.

Since we assume the magnon transport is ballistic, we can use the Landauer approach to investigate the magnetization current $I_m$\cite{Rego1999Fractional, Loss2003MagnetizationTransport, zhang2013topological}, which is given by
\begin{equation}
	I_m(V)=\frac{\textsl{g}\mu_{B}}{8\pi^3}\int_{{\bf k}} v_{x(z)}T({\bf k},V)\left[f_{B,L}(\varepsilon)-f_{B,R}(\varepsilon)\right]d{\bf k},
\end{equation}
where $f_{B,R(L)}(\varepsilon)=1/\{\exp[\beta (\varepsilon+E_{0}\pm \textsl{g}\mu_{B}\Delta B/2)]-1\}$ refers to the Bose distribution functions of the right (left) magnon reservoir, respectively. $\varepsilon$ stands for the energy of the incident magnons with respect to the Weyl point of which the energy is $E_{0}$, ${\bf k}=(k_x,k_y,k_z)$ stands for the momentum of incident magnons and $v_{x(z)}=d\varepsilon/dk_{x(z)}$ is their group velocity in the $x$($z$)-direction. To account for all the possible incident magnons, we must integrate over the whole momentum space. The magnetization current can be written as
\begin{equation}
	\begin{aligned}
		I_m(V)= & I_0\iiint d\varepsilon dk_ydk_{z(x)}T(\varepsilon,k_y,k_{z(x)},V) \left[f_{B,L}(\varepsilon)-f_{B,R}(\varepsilon)\right],
	\end{aligned}
	\label{LB}
\end{equation}
where $I_0=\frac{\textsl{g}\mu_{B}}{8\pi^3}$.

Meanwhile, different from the electronic cases, incident magnons with energy $\varepsilon+E_{0}\in[0,\textsl{g}\mu_{B}\Delta B/2]$ do not transport from $R_L$ to $R_R$\cite{Loss2003MagnetizationTransport}. Thus, it is convenient to keep $\Delta B<(2E_{0}-\Delta\varepsilon)/\textsl{g}\mu_{B}$ to keep all incident magnons from each reservoir can transmitter to another in our discussion.

We assume the bandwidth of the Weyl cone to be $\Delta\varepsilon$, such that the Weyl magnons with energy ranging from $E_0-\Delta\varepsilon/2$ to $E_0+\Delta\varepsilon/2$ (see Fig.\ref{Fig10d}). In addition, we assume there is no other magnon band outside the Weyl cone. As a result, all the magnons contributing to $I_m$ obey the Weyl equation. When $\varepsilon-V<-\Delta\varepsilon/2$ or $\varepsilon-V>\Delta\varepsilon/2$, the transmission probability is vanishingly small since there is no corresponding magnon bands in the barrier region.

Finally, according to previous researches on topological magnonic materials\cite{zhang2013topological,Fransson2016MagnonDirac, Su2017MagnonicWeyl,Su2017ChiralAnomaly,Owerre2018WeylMagnons}, it is suitable to set $\Delta\varepsilon=1$ as an unit, and we choose $E_{0}=1.5$, $\textsl{g}\mu_{B}\Delta B=0.01$ and $\beta=1/k_{B}T=1$ in the following discussions.

\subsection{Isotropic Weyl magnons}
\label{SubSec:CWMagnon}
\begin{figure}[t]
	\centering
	\includegraphics[height=5cm]{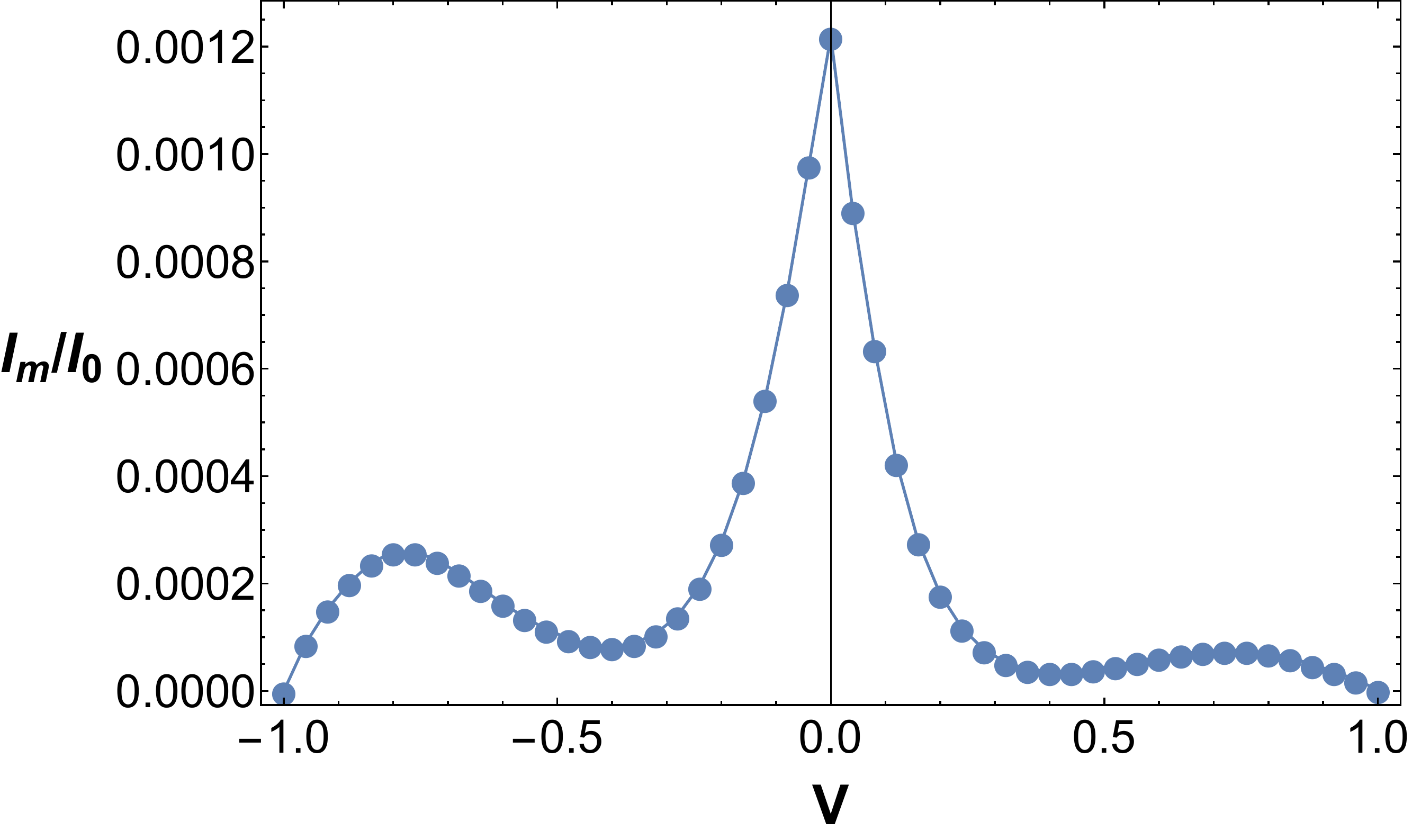}
	\caption{The barrier height dependence of the magnetization current carried by isotropic Weyl magnons, with $E_{0}=1.5$, $\textsl{g}\mu_{B}\Delta B=0.01$, $d=4\pi$, and $\beta=1/k_{B}T=1$. When $B_g$ is positive, it serves as a potential barrier in the magnetic wire; when $B_g$ is negative, it becomes a potential well. The magnetization current vanishes when $|V|>1$ since there are no corresponding magnon bands in the barrier region.}
	\label{Fig11}
\end{figure}

We first consider the isotropic Weyl magnon case. With the transmission probability given in Sec.\ref{SubSec:WMagnon}, the magnetization current influenced by Klein tunneling can be calculated. Fig.\ref{Fig11} shows the barrier height dependence of the magnetization current carried by isotropic Weyl magnons. When $B_g$ is positive, it serves as a potential barrier in the magnetic wire; when $B_g$ is negative, it becomes a potential well instead. The magnetization current $I_m$ vanishes when $V>1$ or $V<-1$ since there are no corresponding magnon bands in the barrier region.

It is convenient to introduce $k_{\perp}=\sqrt{k_y^2+k_{z(x)}^2}$ and $\gamma=\arcsin(k_y/k_{\perp})$ to Eq.(\ref{LB}), which becomes
\begin{equation}
	\begin{aligned}
		I_m(V)= & I_0\int^{\Delta\varepsilon/2}_{-\Delta\varepsilon/2}d\varepsilon\int^{\varepsilon}_{0}dk_{\perp}\int^{2\pi}_{0}d\gamma T(\varepsilon,k_{\perp},\gamma,V) \\&\left[f_{B,L}(\varepsilon)-f_{B,R}(\varepsilon)\right]k_{\perp}.
	\end{aligned}
\end{equation}
Since the Weyl cone is isotropic, $I_m(V)$ in the $z$-direction is the same as in the $x$-direction.

Fig.\ref{Fig11} shows sharp drops when $|V|$ ranges from $0$ to about $0.3$. This is caused by the low transmission probability of incident magnons when the potential barrier or well starts to appear. When $|V|\gtrsim0.3$, the Klein tunneling effect and the Fabry-P\'erot resonances of magnons affect most of the incident magnons, which increases the transmission probability and causes the appearance of the bumps in Fig.\ref{Fig11} when $0.5\lesssim|V|<1$. However, since Bose distribution $f_{B,L}(\varepsilon)-f_{B,R}(\varepsilon)$ is not centered around a certain incident magnon energy, the oscillations of magnons at each energy are canceled out by each other. As a result, $I_m$ does not show a clear oscillation against $V$ but shows bumps. When $|V|>1$, there are not magnon bands in the barrier regions for magnons to get through the potential barrier and $I_m$ vanishes.

\subsection{Weyl magnons with tilted dispersion}
\label{SubSec:CTWMagnon}
We then consider the tilted Weyl cone case. As stated in Sec.\ref{SubSec:TWMagnon}, $uk_x$ in Eq.(\ref{TWHam}) describes the tilting of the Weyl cones. Due to the tilting, the transport properties in the $x$-direction are different from that in the $z$-direction.

When the magnon transport is in the $x$-directions, the incident magnons obey inequality (\ref{ConditionKTWX}). Thus we introduce $k_{\perp}= \sqrt{k_y^2+k_{z}^2}$ and $\gamma=\arcsin(k_z/k_{\perp})$ to Eq.(\ref{LB}). However, different from the isotropic case, $k_{\perp}$ ranges from $0$ to $\frac{|\varepsilon|}{\sqrt{1-u^2}}$. It means that there are more possible incident magnon states when $u$ increases. As a result, Eq.(\ref{LB}) becomes
\begin{equation}
	\begin{aligned}
		I_m(V)= & I_0\int^{\Delta\varepsilon/2}_{-\Delta\varepsilon/2}d\varepsilon\int^{\frac{|\varepsilon|}{\sqrt{1-u^2}}}_{0}dk_{\perp}\int^{2\pi}_{0}d\gamma T(\varepsilon,k_{\perp},\gamma,V) \\&\left[f_{B,L}(\varepsilon)-f_{B,R}(\varepsilon)\right]k_{\perp}.
	\end{aligned}
\end{equation}

\begin{figure}[]
	\centering
	\includegraphics[height=5cm]{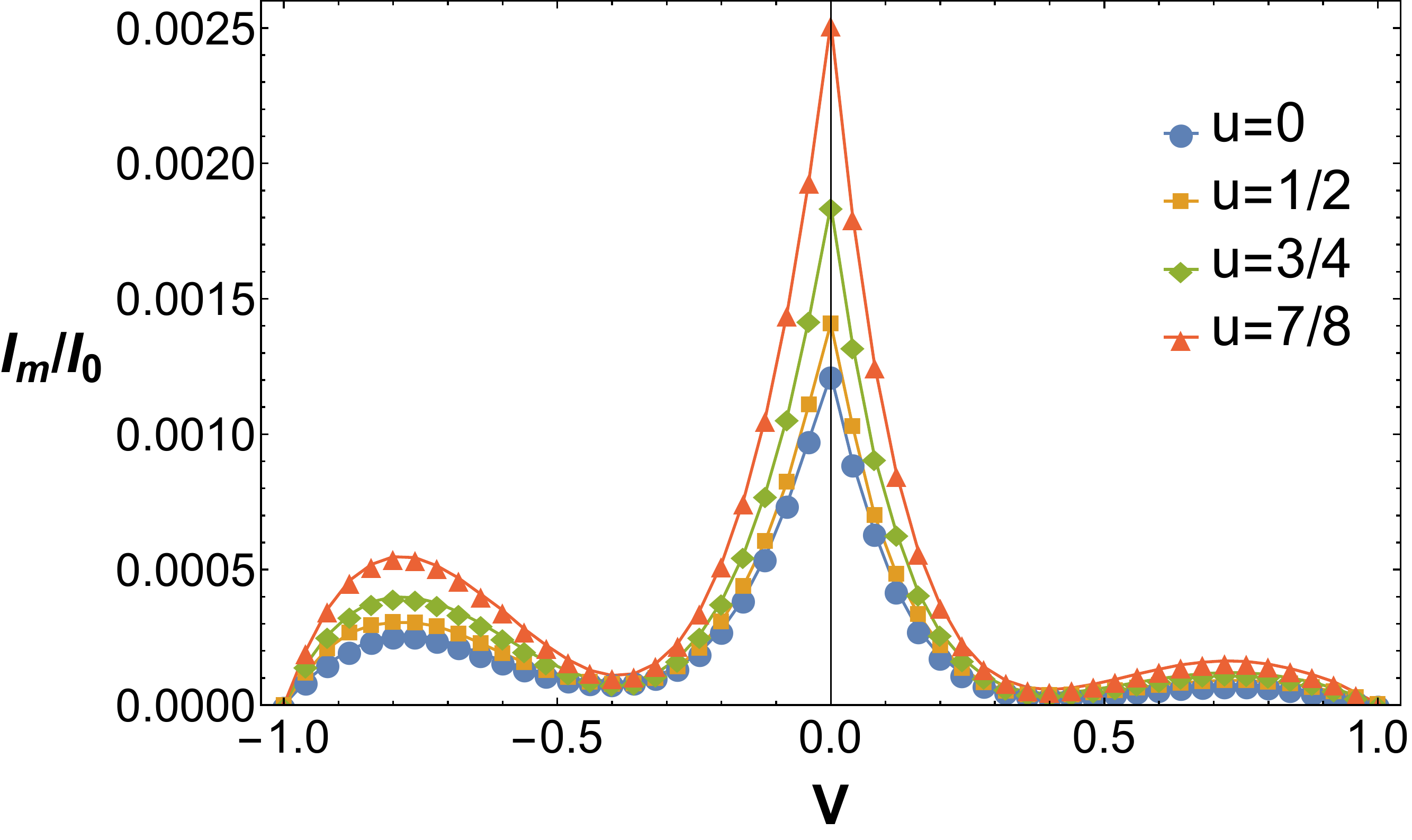}
	\caption{The barrier height dependence of the magnetization current carried by tilted Weyl magnons in the $x$-direction, with the same parameters as in Fig.\ref{Fig11}.}
	\label{Fig12}
\end{figure}

Fig.\ref{Fig12} shows the magnetization current in the $x$-direction carried by tilted Weyl magnons as a function of the barrier height $V$. $I_m$ with a higher $u$ is stronger, due to more incident magnons states.

When the magnons transport in the $z$-direction, the incident magnons obey inequality (\ref{ellispse}) which is an ellipse. Thus, we introduce $k_{\perp}^2=(1-u^2)(k_x+\frac{\varepsilon u}{1-u^2})^2+k_y^2$ and $\gamma=\arcsin(k_y/k_{\perp})$ to Eq.(\ref{LB}). Then, they have
\begin{equation}
	\begin{aligned}
		k_{x} & =\frac{k_{\perp}\cos\gamma}{\sqrt{1-u^2}}-\frac{\varepsilon u}{1-u^2}, \\
		k_{y} & =k_{\perp}\sin\gamma.
	\end{aligned}
\end{equation}
And $k_{\perp}$ ranges from $0$ to $\frac{|\varepsilon|}{\sqrt{1-u^2}}$. Eq.(\ref{LB}) becomes
\begin{equation}
	\begin{aligned}
		I_m(V)= & I_0\int^{\Delta\varepsilon/2}_{-\Delta\varepsilon/2}d\varepsilon\int^{\frac{|\varepsilon|}{\sqrt{1-u^2}}}_{0}dk_{\perp}\int^{2\pi}_{0}d\gamma T(\varepsilon,k_{\perp},\gamma,V) \\&\left[f_{B,L}(\varepsilon)-f_{B,R}(\varepsilon)\right]\frac{k_{\perp}}{\sqrt{1-u^2}}.
	\end{aligned}
\end{equation}

\begin{figure}[]
	\centering
	\includegraphics[height=5cm]{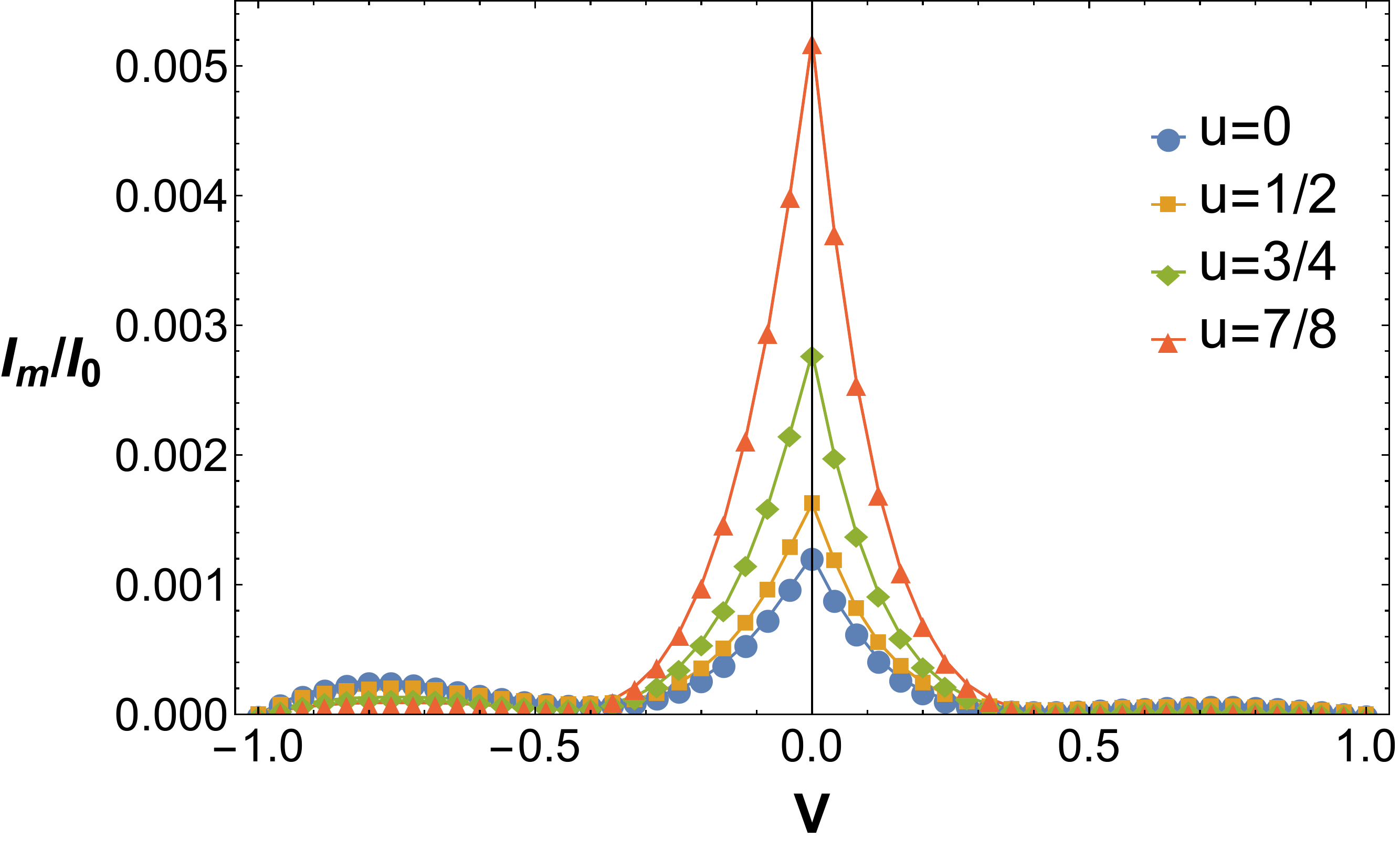}
	\caption{The barrier height dependence of the magnetization current carried by tilted Weyl magnons in the $z$-direction, with the same parameters as in Fig.\ref{Fig11}.}
	\label{Fig13}
\end{figure}

Fig.\ref{Fig13} shows the magnetization current in the $z$-direction carried by tilted Weyl magnons against the barrier height $V$. $I_m$ with higher $u$ is stronger when $|V|\lesssim0.3$, which is also due to more incident magnon states. However, when $|V|\gtrsim0.3$, $I_m$ with a higher $u$ is weaker. It is different from the $x$-direction case. This difference can be understood from the view of the transmission probability. Since the normal incidence in the momentum space deviates from that in the real space as $u$ increases, more and more incident magnons cannot go through the potential barrier and hence have a smaller transmission probability than the untilted case (see Fig.\ref{Fig05}). As a result, the bumps caused by the Klein tunneling effect and Fabry-P\'erot resonances become lower as $u$ increases (see Fig.\ref{Fig13}).

\subsection{Double Weyl magnons}
\label{SubSec:CDWMagnon}
Now we consider the transport of double Weyl magnons. As discussed in Sec.\ref{SubSec:DWMagnon}, the incident magnons obey the Hamiltonian Eq.(\ref{DW}), so the magnon transport properties are anisotropic.

When the magnon transport is in the $x$-direction, the incident magnons satisfy inequality (\ref{ConditionKDWX}). Since $k_y$ is real in the tunneling process and the transmission probability $T(k_x,k_y,k_z)=T(k_x,-k_y,k_z)$, it is convenience to  consider the integrating Eq.(\ref{LB}) with $k_y>0$ and $k_y<0$ respectively. When $k_y>0$, we introduce $k_{\perp}^2=k_y^4+k_z^2$ and $\gamma=\arcsin(k_z/k_{\perp})$ to Eq.(\ref{LB}). Then, they have
\begin{equation}
	\begin{aligned}
		k_{y} & =\sqrt{k_{\perp}\cos\gamma}, \\
		k_{z} & =k_{\perp}\sin\gamma,
	\end{aligned}
\end{equation}
with $k_{\perp}$ ranging from $0$ to $\varepsilon$ and $\gamma$ from $-\pi/2$ to $\pi/2$. The integral Eq.(\ref{LB}) with negative $k_y$ is the same as its positive case. Finally, Eq.(\ref{LB}) becomes
\begin{equation}
	\begin{aligned}
		I_m(V)= & I_0\int^{\Delta\varepsilon/2}_{-\Delta\varepsilon/2}d\varepsilon\int^{|\varepsilon|}_{0} dk_{\perp}\int^{\pi/2}_{-\pi/2}d\gamma T(\varepsilon,k_{\perp},\gamma,V) \\&\left[f_{B,L}(\varepsilon)-f_{B,R}(\varepsilon)\right]\sqrt{\frac{k_{\perp}}{\cos\gamma}}.
	\end{aligned}
\end{equation}

\begin{figure}[]
	\centering
	\includegraphics[height=5cm]{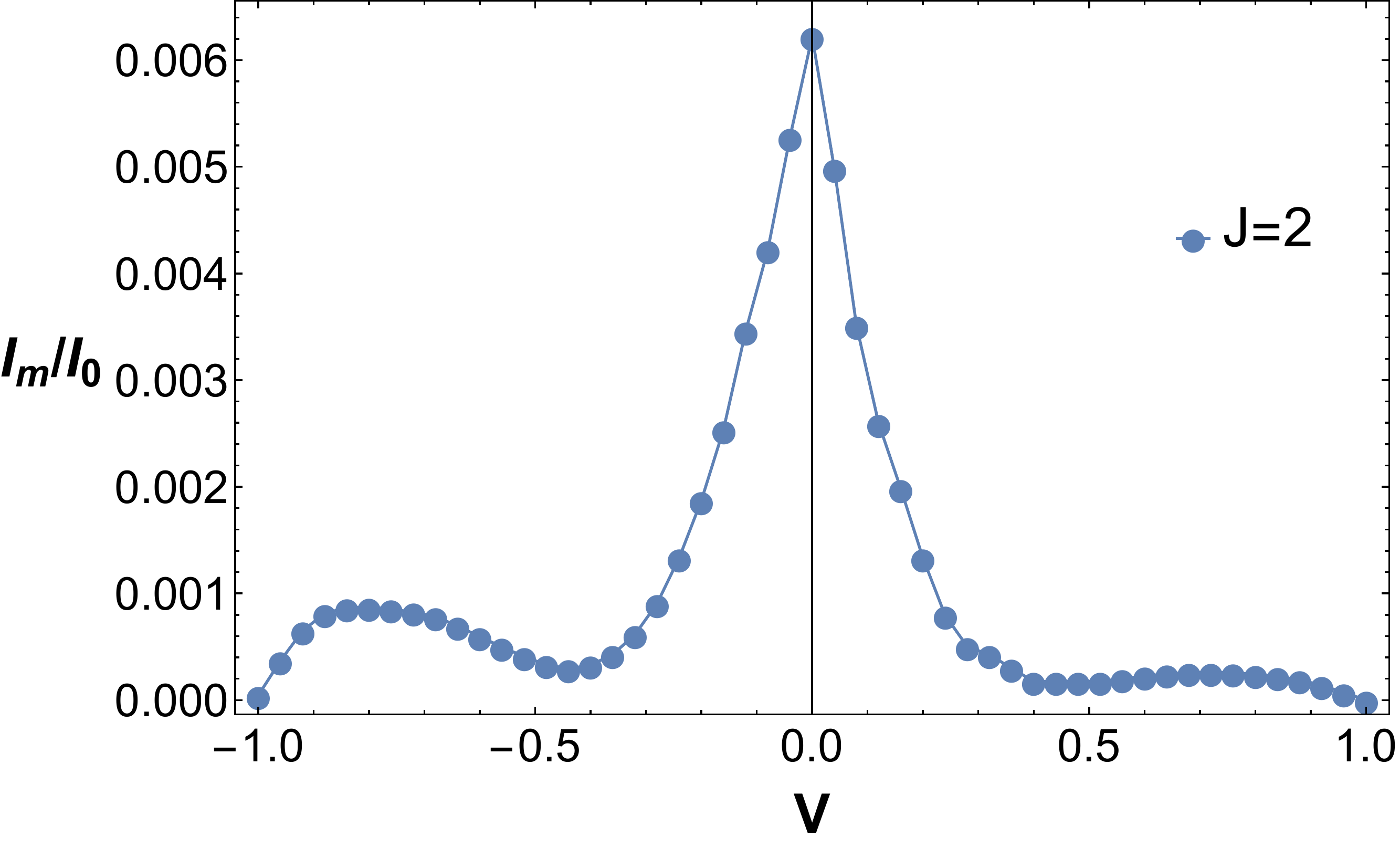}
	\caption{The barrier height dependence of the magnetization current carried by double Weyl magnons in the $x$-direction.}
	\label{Fig14}
\end{figure}

Fig.\ref{Fig14} shows the barrier height dependence of the magnetization current in the $x$-direction carried by double Weyl magnons. Compared with the isotropic Weyl cone case, there are more incident magnon states, resulting in the stronger magnetization current. However, due to the absence of the Klein tunneling effect, the bump in the range $V\in[0.5,1]$ is lower, which is only caused by Fabry-P\'erot resonances. In the potential well case ($V<0$), since we still can observe that transmission probability $T=1$ independent of the well depth, the bump in the range $V\in[-1,-0.5]$ is high and obvious.

When the magnon transport is in the $z$-direction, the incident magnons satisfy inequality (\ref{ConditionKDWZ}). We introduce $k_{\perp}^2=k_x^2+k_y^2$ and $\gamma=\arcsin(k_y/k_{\perp})$ to Eq.(\ref{LB}) with $k_{\perp}$ from $0$ to $\sqrt{|\varepsilon|}$, and $\gamma$ from $0$ to $2\pi$. Finally, Eq.(\ref{LB}) becomes
\begin{equation}
	\begin{aligned}
		I_m(V)= & I_0\int^{\Delta\varepsilon/2}_{-\Delta\varepsilon/2}d\varepsilon\int^{\sqrt{|\varepsilon|}}_{0} dk_{\perp}\int^{2\pi}_{0}d\gamma T(\varepsilon,k_{\perp},\gamma,V) \\&\left[f_{B,L}(\varepsilon)-f_{B,R}(\varepsilon)\right]k_{\perp}.
	\end{aligned}
\end{equation}

\begin{figure}[]
	\centering
	\includegraphics[height=5cm]{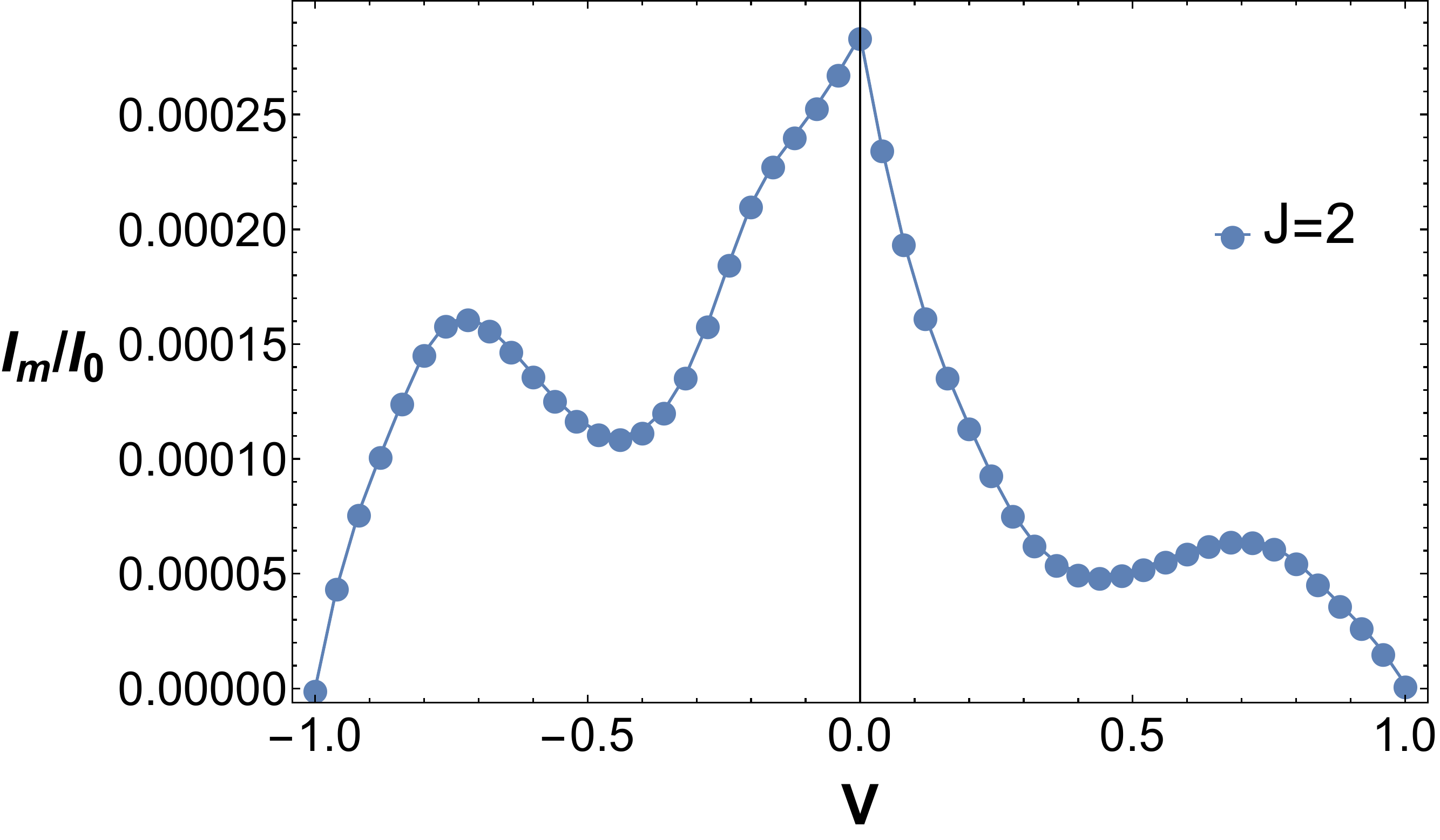}
	\caption{The barrier height dependence of the magnetization current carried by double Weyl magnons in the $z$-direction.}
	\label{Fig15}
\end{figure}

Fig.\ref{Fig15} shows the magnetization current in the $z$-direction carried by double Weyl magnons as a function of the barrier height $V$. As shown in Fig.\ref{Fig09}, the Klein tunneling effect is even stronger than that in the isotropic Weyl cone case. There is a much larger region in the center of Fig.\ref{Fig09} that the transmission probability keeps high with varying $V$. As a result, there are much higher bumps in Fig.\ref{Fig15} than that in Fig.\ref{Fig11}. However, the incident magnon states are quite less than those in the isotropic Weyl cone case. Finally, even though there are stronger fluctuations in the curve of $I_m$, the magnetization current is actually much weaker than that in the isotropic Weyl cone case.

\section{Conclusion}
\label{Sec:CONCLUSION}
In conclusion, we have studied the Klein tunneling of Weyl magnons and its consequences in the magnetization current of a magnetic wire. In particular, we considered a quasi-one-dimensional magnetic wire made by magnonic Weyl materials. The magnetization current in this wire is carried by Weyl magnons, which are driven by the magnon chemical potential difference of two reservoirs at the ends of the wire. We introduced a potential step, either barrier or well, by setting a constant gate magnetic field in the middle of the magnetic wire. With this setting, we investigated Klein tunneling of various kinds of Weyl magnons, including isotropic, tilted and double Weyl magnons. Then we presented the magnetization current curve influenced by the Klein tunneling effect and Fabry-P\'erot resonances of magnon wave functions. As bosonic statistics is in charge, the current-voltage characteristics is rather different from that in the electronic case, which shows clear oscillatory behavior.

In the case of isotropic Weyl magnons, we derived the magnon transmission properties through a potential barrier from the continuity of wave functions. Even the transmission probability $T$ is the same as in the electronic case as they obey the same Hamiltonian, the Bose distribution function makes a difference in the transport properties. Since the difference of Bose distribution $f_{B,L}(\varepsilon)-f_{B,R}(\varepsilon)$ is not centered around a certain incident magnon energy in the above setting, the oscillations induced by Fabry-P\'erot resonances are canceled by each other. As a consequence, the magnetization current $I_m$ does not show a clear oscillation behavior against the gate magnetic field. Instead, the Klein tunneling effect and Fabry-P\'erot resonances cause bumps between $|V|\sim0.5$ and $|V|=1$ (see Fig.\ref{Fig11}).

The tilting of the Weyl cone makes the transmission properties and hence the magnon transport properties anisotropic. Assuming the tilting is in the $x$-direction, we studied these properties in the $x$- and the $z$-direction. When magnons are transported along the magnetic wire in the $x$-direction, a stronger Fabry-P\'erot resonances is observed in the transmission probability figures as the tilting is increased. And the magnetization current becomes stronger since there are more incident magnons states as $u$ increases. When magnon transport happens in the $z$-direction, the normal incidence in momentum space deviates from the point corresponding to that in real space. This behavior makes some of the incident magnons hard or even unable to get through the potential barrier. Finally, although there are more incident magnon states, the two bumps of $I_m$ are still lower than that in the untilted case.

Finally, in the double Weyl magnon case, the Hamiltonian is also anisotropic. When the transport happens in the $x$-direction, the Klein tunneling effect is absent due to the presence of the evanescent waves when $V>0$, but the transmission probability $T=1$ independent of the depth of the potential well when $V<0$. Consequently, we observed that the curve of the magnetization current keeps a high bump when $V<0$ while a lower one when $V>0$. When the magnetization current is in the $z$-direction, it is much weaker than the isotropic Weyl cone case, due to less incident magnon states. However, the Klein tunneling effect is even stronger, which can be seen from the expanding of the areas that $T=1$ independent of the potential barrier. This difference causes the higher bump in the curve of the magnetization current.

\begin{acknowledgments}
	L.W. thanks S.-K. Jian for useful discussions. This work was supported by NKRDPC-2017YFA0206203 and 2018YFA0306001, NSFC-11974432, NSFC-11574404, National Supercomputer Center in Guangzhou and Leading Talent Program of Guangdong Special Projects.
\end{acknowledgments}

\bibliography{ref}
\bibliographystyle{apsrev4-1}

\end{document}